%% file: main.tex
\documentclass[letterpaper,twocolumn,10pt]{article}
\usepackage{usenix}

\usepackage{tikz}
\usepackage{amsmath}
\usepackage{amsfonts}
\usepackage{algorithmicx}
\usepackage{algpseudocode}
\usepackage{algorithm}
\usepackage{graphicx}
\usepackage{subcaption}
\usepackage{float}
\usepackage{xspace}
\usepackage{xurl}
\usepackage{subcaption}
\usepackage{enumitem}
\usepackage[available]{usenixbadges}
\DeclareMathOperator*{\lat}{Lat}
\DeclareMathOperator*{\size}{Size}

\newcommand{\sys}{Cocoon\xspace}
\newcommand{\sysdlrm}{Cocoon-Emb\xspace}

\newcommand{\otfgpu}{GPU-GEMV\xspace}
\newcommand{\otfcpu}{CPU-GEMV\xspace}
\newcommand{\Mod}[1]{\ (\mathrm{mod}\ #1)}
\newcommand{\latX}{\lat\nolimits_{\text{gemv-}X}}
\newcommand{\latgpu}{\lat\nolimits_{\text{gemv-}G}}
\newcommand{\latcpu}{\lat\nolimits_{\text{gemv-}C}}
\newcommand{\latnmp}{\lat\nolimits_{\text{gemv-}N}}
\newcommand{\munit}{m_u}
\newcommand{\lattraingpu}{\lat\nolimits_{\text{train-}G}}

\newcommand{\memtotalX}{Mem_{\text{total-}X}}
\newcommand{\memtotalgpu}{Mem_{\text{total-}G}}
\newcommand{\memtotalcpu}{Mem_{\text{total-}C}}
\newcommand{\memtotalnmp}{Mem_{\text{total-}N}}

\newcommand{\memtrainX}{Mem_{\text{train-}X}}
\newcommand{\memtraingpu}{Mem_{\text{train-}G}}
\newcommand{\memtraincpu}{Mem_{\text{train-}C}}

\begin{document}

\date{}

\title{\Large Cocoon: A System Architecture for Differentially Private \\ Training with Correlated Noises}

\author{
{\rm Donghwan Kim}$^1$,
{\rm Xin Gu}$^1$,
{\rm Jinho Baek}$^2$,
{\rm Timothy Lo}$^2$,
{\rm Younghoon Min}$^2$, \\
{\rm Kwangsik Shin}$^2$, 
{\rm Jongryool Kim}$^2$,
{\rm Jongse Park}$^3$,
{\rm Kiwan Maeng}$^1$
\\[0.5em]
$^1$The Pennsylvania State University \\
$^2$SK hynix Inc.  
$^3$KAIST \\
\\[0.5em]
}

\maketitle
\thispagestyle{empty}   
\vspace{5em}

\begin{abstract}
\input{Chapters/0_abstract}
\end{abstract}

\pagestyle{empty}
\input{Chapters/1_introduction}
\input{Chapters/2_background}
\input{Chapters/3_characterization}
\input{Chapters/4_Coalescing}
\input{Chapters/5_NMP}

\input{Chapters/7_evaluation}
\input{Chapters/8_related}
\input{Chapters/9_discussion}
\input{Chapters/10_conclusion}

\section*{Acknowledgments}
This work was partly supported by the U.S. National Science Foundation under award No. CNS-2349610 and CCF-2529883. Any opinions, findings, and conclusions or recommendations expressed in this material are those of the author(s) and do not necessarily reflect the views of the National Science Foundation. 

\bibliographystyle{plain}
\bibliography{refs}

\input{Chapters/osdi22_ae_appendix_template}

\end{document}

%% file: Chapters/0_abstract.tex
Machine learning (ML) models memorize and leak training data, causing serious privacy issues to data owners.
Training algorithms with differential privacy (DP) have been gaining attention as a solution.
However, these algorithms add noise at each training iteration and degrade accuracy, limiting their real-world adoption.
To improve accuracy, a new family of approaches adds carefully designed \emph{correlated noises}, so that noises cancel out each other across iterations.
We performed an extensive characterization study of these new mechanisms
and show they incur non-negligible overheads when the model is relatively large or uses large embedding tables compared to the hardware capacity.
Motivated by the analysis, we propose \sys, a framework for efficient training with correlated noises.
\sys stores and processes the large noise history across CPU, GPU, and memory extension module, introduces optimizations for sparse embedding tables, and leverages to-be-commercialized near-memory processing (NMP) devices.
On a real system with an FPGA-based NMP device prototype, \sys improves the performance by 1.23--10.82$\times$.

%% file: Chapters/1_introduction.tex
\section{Introduction}
\label{sec:introduction}

Machine learning (ML) models are vulnerable to memorization and leakage of their training data---sometimes by accident, or when model is subjected to an adversarial attack. 
%
This poses a serious \emph{privacy risk} because ML models are often trained with sensitive user data (\emph{e.g.}, medical data~\cite{zhang2022shifting}, audio~\cite{michaely2017keyword} or keyboard inputs~\cite{rahul_2013_keyboardprivacy, gboard_ftrl_1}, personal chat logs~\cite{yu_2024_privacy_preserving_instruction_alignment}, \emph{etc}.).
The threat is not hypothetical but imminent: a recent study demonstrated the feasibility of attacking the popular ChatGPT~\cite{nasr_2025_chatgpt_attack} to extract its training data that contains sensitive information. Similar attacks have been shown to be possible for other models~\cite{carlini_2023_diffusion_attack, carlini_2021_gpt2_attack, fredrikson_2015_model_inversion, kariyappa_2023_cpa}.
%
%
The privacy risks can deter user participation in data collection and training, degrading the quality of real-world ML-based services~\cite{facebook_idfs}.

Differential privacy (DP)~\cite{dwork2006differential} is one of the most principled and widely adopted approaches for mitigating such privacy risks.
DP theoretically bounds the information leakage of private data through its statistics. 
In the context of ML training, this translates to a mathematical bound on the probability that an adversary can successfully extract information of the training data from the trained model~\cite{kairouz_2015_composition, nasr_2021_adversary, guo_2022_bounding, guo_2023_fano, hayes_2023_bounding}.
DP is not a premature research concept; already deployed in several real-world applications, including the US Census~\cite{census_dp}, Apple's device analytics~\cite{apple_dp_deployment, apple_dp_problem}, and Google Chrome's behavior telemetry~\cite{chrome_rappor}.
The use of DP for ML training is also gaining increased attention. For instance, Google publicly announced its use of DP for training all of their current and future smart keyboard (Gboard) models~\cite{gboard_ftrl_1, gboard_ftrl2}, and also recently released VaultGemma~\cite{sinha_2025_vaultgemma}, an LLM that was trained with DP from scratch.
Microsoft has also limitedly deployed a DP-trained model on their SwiftKey keyboard~\cite{swiftkey_dp}.

%
%
%
Most DP training algorithms add Gaussian noise to the gradient at each training iteration to provide a theoretical privacy guarantee~\cite{abadi_2016_deep, de_2022_unlocking}.
%
This Gaussian noise accumulates over training and significantly degrades the model accuracy, which is the primary barrier to DP's widespread adoption.
%
%
%
To counteract the accuracy issue, a recent line of works~\cite{kairouz_2021_practical, choo_2023_multi,choquette-choo_2025_near_exact,choquette_2023_correlated,mcmahan_2024_hassle,neurips_2023_amplified,mckenna_2025_scaling,mckenna_2024_scaling, pillutla2025correlated, jiaojiao_2025_locally} proposed to use noises that are \textbf{\emph{correlated}} across iterations, allowing noises added in later iterations can partially cancel out earlier noises.
%
%
%
{\emph{Correlated noise mechanisms}} have demonstrated both theoretical~\cite{correlated_noise_theory, choquette_2023_correlated} and empirical~\cite{kairouz_2021_practical, choo_2023_multi,choquette-choo_2025_near_exact,choquette_2023_correlated,mcmahan_2024_hassle,neurips_2023_amplified,mckenna_2025_scaling,mckenna_2024_scaling, pillutla2025correlated, jiaojiao_2025_locally} benefits over simply adding Gaussian noise. The mechanism has already been successfully deployed in real-world products~\cite{gboard_ftrl_1, gboard_ftrl2}, and adopted into popular ML privacy libraries~\cite{filip_2025_pfl, tensorflow_privacy}.
Their rapid adoption send a strong signal of their importance.
%

Despite their growing popularity, 
the system implications of correlated noise have not been thoroughly studied.
Existing literature mostly focused on accuracy rather than system efficiency, training relatively small models with a large number of TPUs/GPUs (\emph{e.g.}, using 1024 machines to train a billion-parameter model~\cite{mckenna_2024_scaling}). In these \emph{resource-abundant} environments, noise-related overheads are small.
%
However, our study reveals that correlated noise incurs significant system overheads, including a memory footprint that exceeds the parameter size by over an order of magnitude, and causing up to 14.49$\times$ runtime slowdown. Such costs are particularly severe for \emph{economical} setups with fewer TPUs/GPUs, which is a configuration frequently studied to enable large model training on small and mid-sized entities ~\cite{bu_2023_zero, ren_2021_zero_offload, rajbhandari_2021_zero3, rajbhandari_2020_zero12, liang_2024_dpzero}.
%
%
%
This overhead arises because generating correlated noises involve storing and processing noise history from past iterations, which can scale up to hundreds of gigabytes or even terabytes.
It becomes especially problematic for (1) models with large embedding tables and (2) large-scale, billion-parameter models (\emph{e.g.}, LLMs). 

We introduce \emph{\sys}, a 
framework for efficient at-scale DP training with correlated noises.
\sys splits the large noise history between CPU, GPU,
and (if present) additional memory extension module, 
and processes them in a distributed fashion to maximize performance.
To mitigate managing noise history for large embedding tables, \sys pre-computes and stores the noise for embedding tables in a compact format by exploiting their gradient sparsity (Section~\ref{sec:Cocoon-DLRM}).
Furthermore, when part of the noise history is offloaded to a memory extension module, we show that leveraging emerging near-memory processing (NMP) solutions~\cite{samsung_nmp, cmm-ax} can greatly improve performance and reduce the cost of deployment (Section~\ref{sec:Cocoon-nmp}).
%
Our evaluation on a real system, with an FPGA-based industry prototype NMP device, showed 1.23--10.82$\times$ speedup on various models and training scenarios.
We summarize our contributions:
\begin{itemize}[noitemsep, leftmargin=*]
    \item We present an extensive system characterization study of emerging DP training methods that use correlated noises.
    Our analysis reveals that, on an economical setup where the number of GPUs is moderate, correlated noise generation can introduce non-negligible slowdown.
    \item We introduce \sys, a highly-optimized, PyTorch-based DP training library that uses correlated noises.
    \sys splits and processes noise history on CPU, GPU, and (if present) emerging NMP devices.
    \item \sys introduces a noise pre-computing and coalescing optimization that accelerates training with large embedding tables by 2.33--10.82$\times$. With to-be-commonly-deployed NMP devices, \sys can achieve 1.23--2.32$\times$ speedup over the baseline when training large models, which we demonstrate with a real industry prototype.
\end{itemize}

%% file: Chapters/2_background.tex
\section{Background and Motivation}
\label{sec:Background}
\subsection{Differentially Private Training Algorithms}
\label{sec:BackgroundDP}

\paragraph{Differential privacy.}
Differential privacy~\cite{dwork2006differential} (DP) provides a robust guarantee that the outcome of a randomized mechanism does not change significantly when a single entry in the input dataset changes.
When applied to ML training, DP ensures that the final trained model does not depend significantly on a single sample (or user~\cite{chua_2024_mind, charles_2024_fine_tuning}) in the training corpus, which effectively limits an adversary's capabilities.
For example, DP parameters can theoretically bound the attack success rate for adversaries attempting to infer whether a user participated during training~\cite{kairouz_2015_composition, nasr_2021_adversary} or trying to directly reconstruct the original training data~\cite{guo_2022_bounding, guo_2023_fano, hayes_2023_bounding} from a DP-trained model.
A model trained with DP can be used for multiple inferences over time, and it will not leak additional information beyond the privacy cost paid upfront during training.
This is in contrast with inference-time DP techniques~\cite{maaten_2020_tradeoffs, tang_2024_privacypreserving,koga_2025_privacypreserving,ginart_2022_submix}, which leak more information about the corpus with more inferences~\cite{pomomerva_2023_DPfy, maaten_2020_tradeoffs}.

%
%

\begin{figure}[t]
    \centering
    \includegraphics[width=0.47\textwidth]{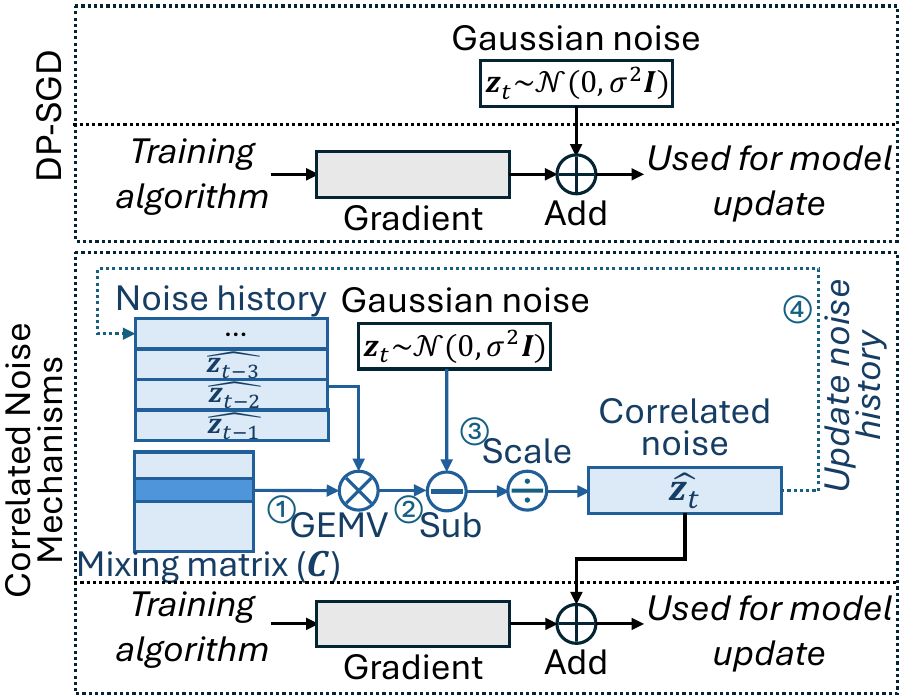}
    \caption{Independent Gaussian vs. correlated noise. 
    }
    \label{fig:DPSGD_vs_FTRL}
\end{figure}


\paragraph{Training with Gaussian noise.}
The most common method to train a model with DP is using DP-SGD~\cite{abadi_2016_deep} or its variants.
DP-SGD introduces three key modifications to regular SGD. 
First, the batch in each iteration is formed through random sampling with replacement.
Second, unlike SGD, which calculates an average gradient across the batch, DP-SGD first calculates per-sample gradients, scales (clips) each one individually, and then averages the scaled gradients.
Finally, an independently sampled Gaussian noise is added to the averaged gradient before it is applied to update the model~\cite{abadi_2016_deep}.
While these steps provide stronger privacy guarantees, DP-SGD significantly compromises the trained model's accuracy.

Several follow-up works have introduced optimizations to improve the model accuracy over the DP-SGD algorithm. Most of this research still relies  on \emph{independent Gaussian noise} and focuses on optimizing other components of DP-SGD (e.g., sampling~\cite{chua_2024_scalable}, clipping~\cite{dp_llm_6, dope_sgd}, hyperparameter~\cite{de_2022_unlocking}, feature engineering~\cite{handcrafted_dp}, post-processing after noise addition~\cite{gu2025choosing, zhang2024disk}, optimizer~\cite{ganesh_2025_dp_optimizers, tang2023dpadam, chooi2025dpadamw, tang2024dpadambc, yang2025dpadamac}, and architecture~\cite{papernot2021tempered, de_2022_unlocking}).
These optimizations still suffer from accuracy degradation due to Gaussian noise.
Figure~\ref{fig:DPSGD_vs_FTRL} (top) highlights the independent Gaussian noise added to each gradient in DP-SGD.

\paragraph{Correlated noise mechanisms.}
Instead of adding independent Gaussian noise, a recent line of works~\cite{kairouz_2021_practical, choo_2023_multi,choquette-choo_2025_near_exact,choquette_2023_correlated,mcmahan_2024_hassle,neurips_2023_amplified,mckenna_2025_scaling,mckenna_2024_scaling, pillutla2025correlated} proposed to add noise that is \emph{correlated} across iterations.
%
%
When generating correlated noises at iteration $t$, these mechanisms mix noises that were used in the previous $\hat{b}-1$ iterations (\emph{i.e.}, noise history), where $\hat{b}$ is called the \emph{band size}.

Mathematically, correlated noises are generated as follows.
For a model with $m$ trainable parameters trained through $n$ training iterations,
let $\mathbf{z}_t \in \mathbb{R}^m$ be a Gaussian noise sampled at iteration $t$. Each noise is as large as the trainable parameters ($m$), and training a large model requires using a noise that is as large.
%
%
%
For a specific \emph{mixing matrix} $\mathbf{C} \in \mathbb{R}^{n \times n}$,
the correlated noise at iteration $t$, $\hat{\mathbf{z}}_t \in \mathbb{R}^m$, can be calculated as:
\begin{equation}
    \hat{\mathbf{z}}_t = (\mathbf{z}_t - \sum_{\tau=1}^{\min (t, \hat{b}-1)} \mathbf{C}[t, t -\tau] \hat{\mathbf{z}}_{t-\tau})/\mathbf{C}[t, t]
    \label{eq:bandedmf}.
\end{equation}

Figure~\ref{fig:DPSGD_vs_FTRL} (bottom) visualizes how $\hat{\mathbf{z}}_t$ is calculated.
First, a \textcircled{1} \emph{weighted average} of the $\hat{b}-1$ previous noises is performed. The result is \textcircled{2} subtracted from a newly-sampled Gaussian noise ($\mathbf{z}_t$) and \textcircled{3} properly rescaled, which becomes the new correlated noise ($\hat{\mathbf{z}}_t$).
As in DP-SGD, $\hat{\mathbf{z}}_t$ is added to the gradient, and the noisy gradient is used to update the model. At the same time, \textcircled{4} $\hat{\mathbf{z}}_t$ is saved to the noise history, so that it can be used to generate future noises.

The weighting/rescaling factors at $t$-th iteration are defined by the $t$-th row of the mixing matrix $\mathbf{C}$, and the weighted averaging can be done through a \emph{matrix-vector multiplication (GEMV)} between the stacked noise history (matrix of size $(\hat{b}-1)\times m$) and the $t$-th row of $\mathbf{C}$ (each row is of size $n$ but only has $\hat{b}-1$ nonzero elements).
$\mathbf{C}$ should be carefully designed to guarantee DP, and different prior works developed different designs of  $\mathbf{C}$~\cite{neurips_2023_amplified,mcmahan_2024_hassle,kairouz_2021_practical}. 
When $\hat{b}=1$ and $\mathbf{C}=\mathbf{I}$ (an identity matrix), this reduces to independent Gaussian noise.


\paragraph{Generality and importance of correlated noise.}
As Figure~\ref{fig:DPSGD_vs_FTRL} shows, correlated noise can almost serve as a drop-in replacement for independent Gaussian noise\footnote{When using correlated noise, it puts a restriction on what data sampling method can be used, but other parts of the training do not need to change.}.
%
Thus, making the correlated noise generation efficient is likely to benefit a wide range of DP training algorithms that currently use independent Gaussian noise.
%
%
Several recent works have demonstrated that correlated noise has theoretical~\cite{correlated_noise_theory, choquette_2023_correlated} and empirical~\cite{kairouz_2021_practical, choo_2023_multi,choquette-choo_2025_near_exact,choquette_2023_correlated,mcmahan_2024_hassle,neurips_2023_amplified,mckenna_2025_scaling,mckenna_2024_scaling, pillutla2025correlated, jiaojiao_2025_locally} 
benefits over independent Gaussian noise.
Google recently adopted correlated noise in its production (Gboard training~\cite{gboard_ftrl_1, gboard_ftrl2}) and has been exploring using it for larger models~\cite{mckenna_2024_scaling, pillutla2025correlated}, which strongly signals its practical importance.
A recent 212-page report~\cite{pillutla2025correlated} summarizes the latest efforts around correlated noise.
Its generality and importance motivate our study.



\subsection{Training of Embedding Tables}
\label{subsubsec:BackgroundDLRM}
An embedding table is a trainable data structure that converts categorical features into a dense vector.
It is commonly used in deep learning recommendation models (DLRMs) or large language models (LLMs).
%
%
%
Especially in DLRMs, embedding tables are very tall and dominate the model size~\cite{maxim_2019_dlrm, gupta_2020_deeprecsys, lim_2024_lazydp}.
%
During each training iteration, only rows (entries) of the table corresponding to the present feature values of the input are used, which leads to several unique behaviors.
%
First, the training speed grows sub-linearly with its size compared to other models, because only a tiny subset is used even when the entire table is large.
Second, unused entries in each iteration have zero gradients. Still, DP training requires noise addition to these zero gradients for privacy~\cite{Ning_2022_eana, lim_2024_lazydp}.
Third, only the entries accessed in each iteration contribute to the gradient calculation at that iteration.
%
%
As we will show later, the first characteristic leads to a unique overhead when using correlated noises, and the latter two will be leveraged by \sys's optimizations (Section~\ref{sec:Cocoon-DLRM}).
%

\subsection{CXL and Near-memory Processing (NMP)}
\label{sec:BackgroundCXLNMP}

Compute express link (CXL) is an open industry interconnect standard based on PCIe, which allows high-speed loads and stores for memory expansion modules plugged into the PCIe slot.
Memory expansion module connected through CXL, or \emph{CXL memory}, has emerged as a way to expand memory capacity.
We study the use of such CXL memory to expand the device memory capacity when the noise history is too large to be hosted on main memory.

While CXL memory provides larger memory, data access through CXL is much slower than the main memory of the CPU.
Several recent works~\cite{park_2024_lpddr, ko2025cosmos, gu_2025_cent, sim2022computational, ke2020recnmp, yun_2024_clay} have proposed to put additional compute units in the CXL controller to reduce data movement through PCIe, and some have made it into (near-)real products~\cite{cmm-ax, samsung_nmp}.
In Section~\ref{sec:Cocoon-nmp}, we show how such emerging \emph{near-memory processing (NMP) CXL memory} products can accelerate \sys.

%% file: Chapters/3_characterization.tex
\section{Characterization of Correlated Noise}
\label{sec:Characterization}

\begin{figure}[t]
  \centering
  \includegraphics[width=0.98\columnwidth]{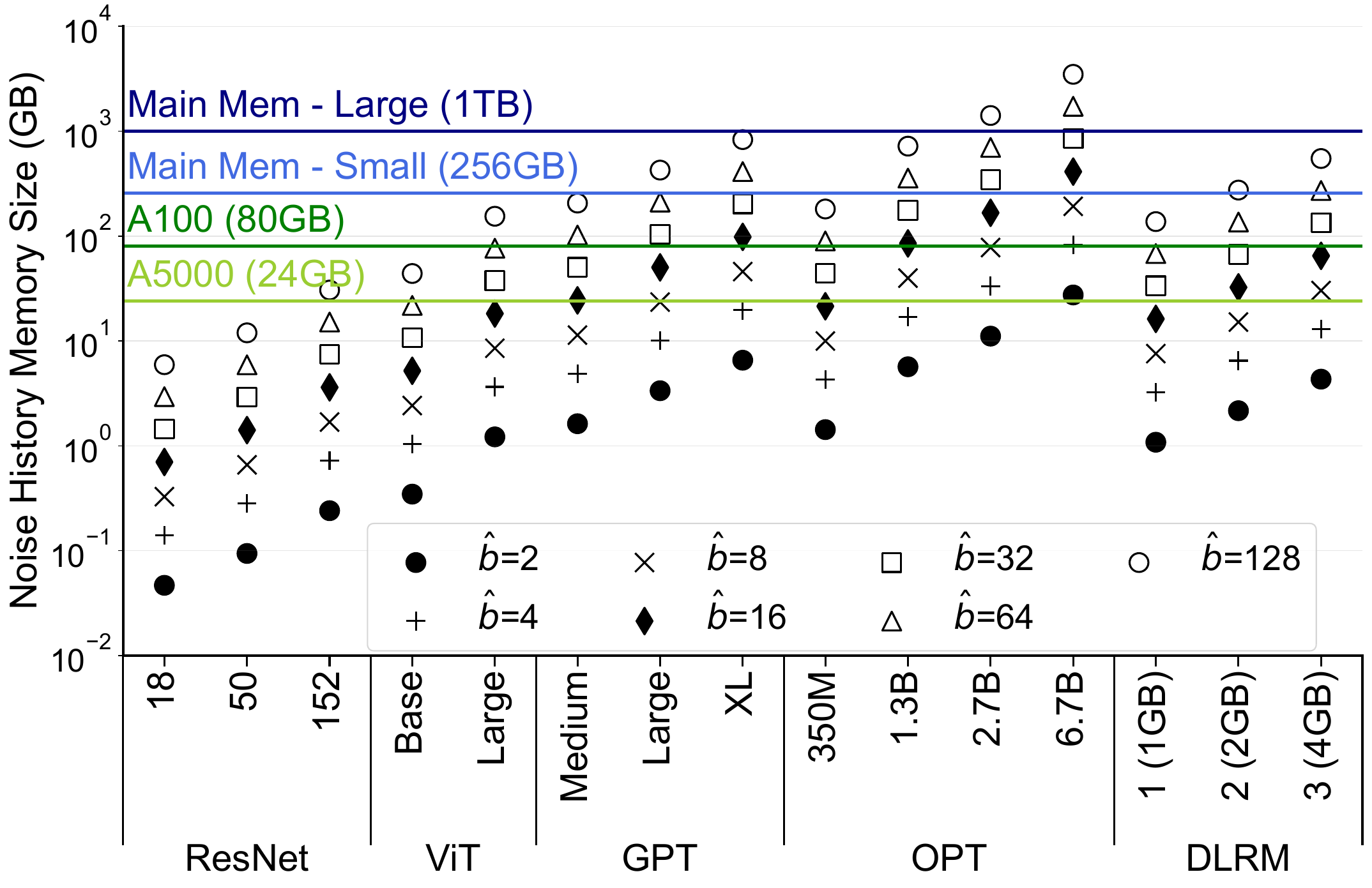}
  \caption{Noise history size of various ML models and $\hat{b}$. 
  }
  \label{fig:NoiseHistorySize}
\end{figure}


To study the system overheads of correlated noise, we trained various models with independent Gaussian noise and correlated noise.
We used a correlated noise mechanism called BandMF~\cite{neurips_2023_amplified}. Our study can be generalized to other mechanisms, because other mechanisms only differ in the derivation of mixing matrix $\mathbf{C}$ (Section~\ref{sec:BackgroundDP}), and are computationally equivalent.
Our study highlights that correlated noise mechanisms experience non-negligible memory (Section~\ref{subsec:Characterize-MemOverhead}) and compute (Section~\ref{subsec:Characterize-CompOverhead}) overheads.
%

%
For the study, we trained popular models from prior DP training literature on a dual-socket Intel Xeon Gold 6330 CPU with 256GB DRAM and 8x NVIDIA A5000 GPUs, which represents an economical setup of a small/mid-sized entity.
The models we trained include: ResNet~\cite{cvpr-2016-resnet}, ViT~\cite{dosovitskiy_2020_image_vit}, LLMs (GPT~\cite{gpt} and OPT~\cite{zhang_2022_opt}), and DLRM~\cite{maxim_2019_dlrm}.
Measurements were done on our custom DP training code with correlated noise support, which we built as part of our \sys library (Section~\ref{sec:Cocoon}).
%
We only highlight the most interesting results.

\subsection{Memory Overhead}
\label{subsec:Characterize-MemOverhead}

Generating correlated noise involves storing  $\hat{b}-1$ past noises, each of which is as large as the number of trainable parameters ($m$).
This introduces several capacity and performance issues.
%

\paragraph{Capacity issue.}
Figure~\ref{fig:NoiseHistorySize} summarizes the memory footprint of the noise history for various models and $\hat{b}$, along with common GPU memory and main memory (CPU DRAM) sizes.
%
The footprint frequently exceeds the GPU/CPU memory, and the noise history must be offloaded to main memory or secondary memory (\emph{e.g.}, CXL memory or SSD) to run the training at all in these cases.
Performance implications of these fallbacks are discussed in Section~\ref{subsec:Characterize-CompOverhead}.
%
%
Some works~\cite{mckenna_2024_scaling, pillutla2025correlated} found such offloading to be unnecessary, but this was because they had enough GPUs/TPUs, unlike our economical setup.
%
%
%
%
%

\emph{Takeaway 1: The noise history can become larger than the aggregate GPU/TPU memory or even the main memory.}

\begin{figure}[t]
    \centering
    \includegraphics[width=0.45\textwidth]{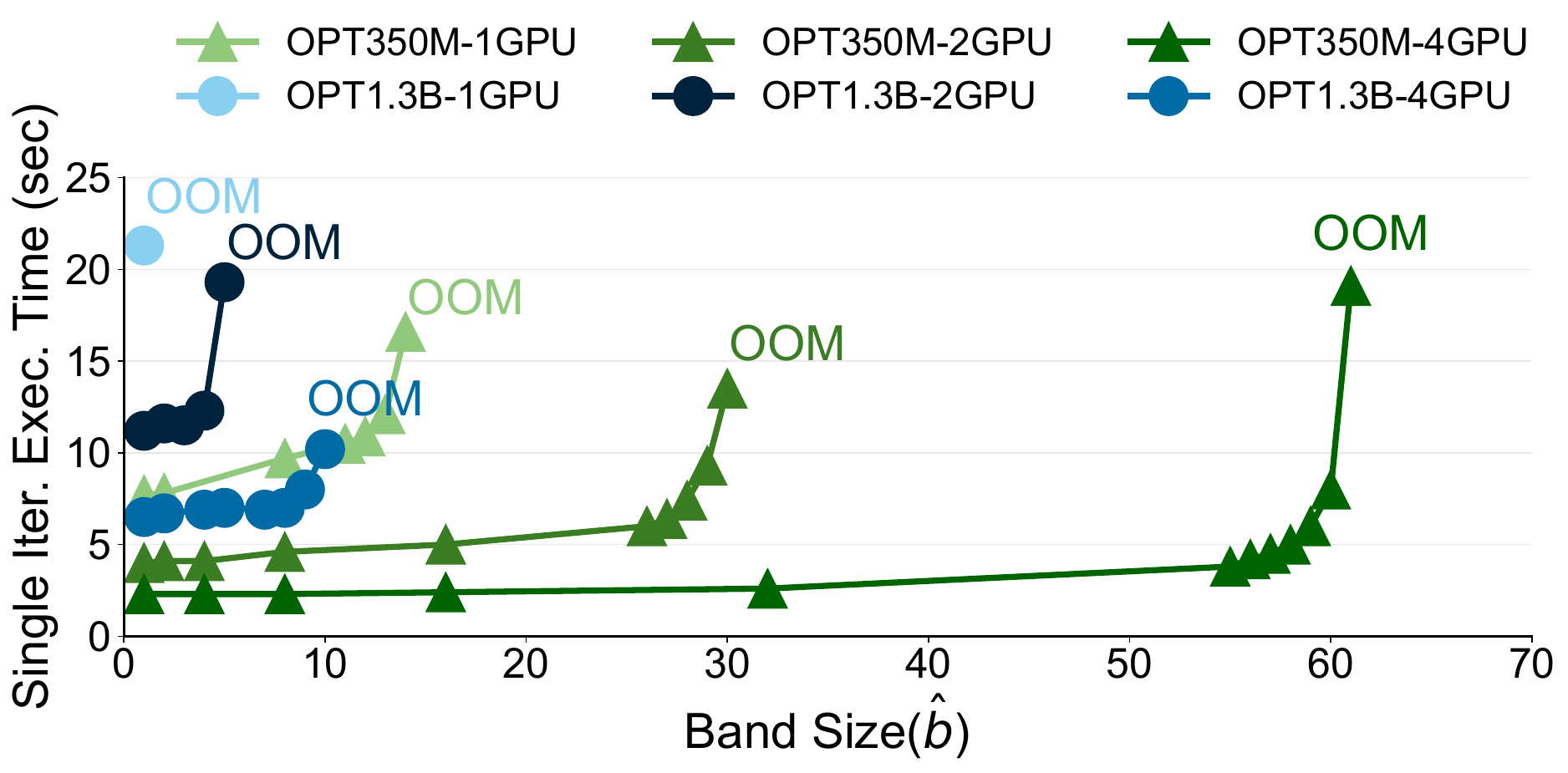}
    \caption{Training time of OPT~\cite{zhang_2022_opt} on 1--4 A5000 GPUs.
    }
    \label{fig:Band-Latency}
\end{figure}

\paragraph{Performance issue.}
Even when the entire noise history fits into the GPU, the performance can degrade if too little memory is left for training.
Figure~\ref{fig:Band-Latency} shows the training latency of two OPT models~\cite{zhang_2022_opt}, where different lines correspond to different models and GPU counts.
It can be seen that the training time increases with $\hat{b}$ for all setups, until an out-of-memory (OOM) error is triggered. 
%
This is because if too little memory is left, training must use smaller microbatches, underutilizing the GPU~\cite{wang2024flashdp}.
%
%

%

\emph{Takeaway 2: Even when storing the entire noise on the GPU is possible, it may degrade the training performance.
}


\paragraph{Why not re-generate noises?}
Instead of storing past noises, prior work~\cite{kairouz_2021_practical} considered only storing the seed and re-generating noises on every iteration.
However, doing so requires re-generating all the noises from the beginning ($\hat{\mathbf{z}}_1,...,\hat{\mathbf{z}}_t$) and not just the past $\hat{b}-1$ noises, because each noise generation recursively requires its past $\hat{b}-1$ noises.
This incurs $O(n^2)$ overhead for $n$ training iterations, which becomes too large unless $n$ is very small.
Another prior work~\cite{pillutla2025correlated} similarly observed that this approach scales poorly with $n$.

\subsection{Compute Overhead}
\label{subsec:Characterize-CompOverhead}

The weighted averaging of prior noises (GEMV between the noise history and the mixing vector) also incurs a computational overhead.
%
%
%
%
%
We mainly consider economical setups where part of the noise history is stored in main/secondary memory.

For the parts hosted outside the GPU, we studied two options for processing them:
(1) \emph{\otfgpu} performs GEMV only on the GPU.
While GEMV is fast on a GPU, this design requires additional data transfer from the main/secondary memory to GPU through the slow PCIe bus.
(2) \emph{\otfcpu} performs GEMV on the CPU for the subset of the noise history stored in main/secondary memory, and only sends the result to the GPU.
While CPU's GEMV is slower, this design enjoys the higher bandwidth between the CPU and main memory, compared to the PCIe bus.
Also, CPU-side GEMV can happen \emph{in parallel} with GPU-side training and can be partially or completely hidden.

\begin{figure}[t]
  \centering
  \includegraphics[width=0.48\textwidth]{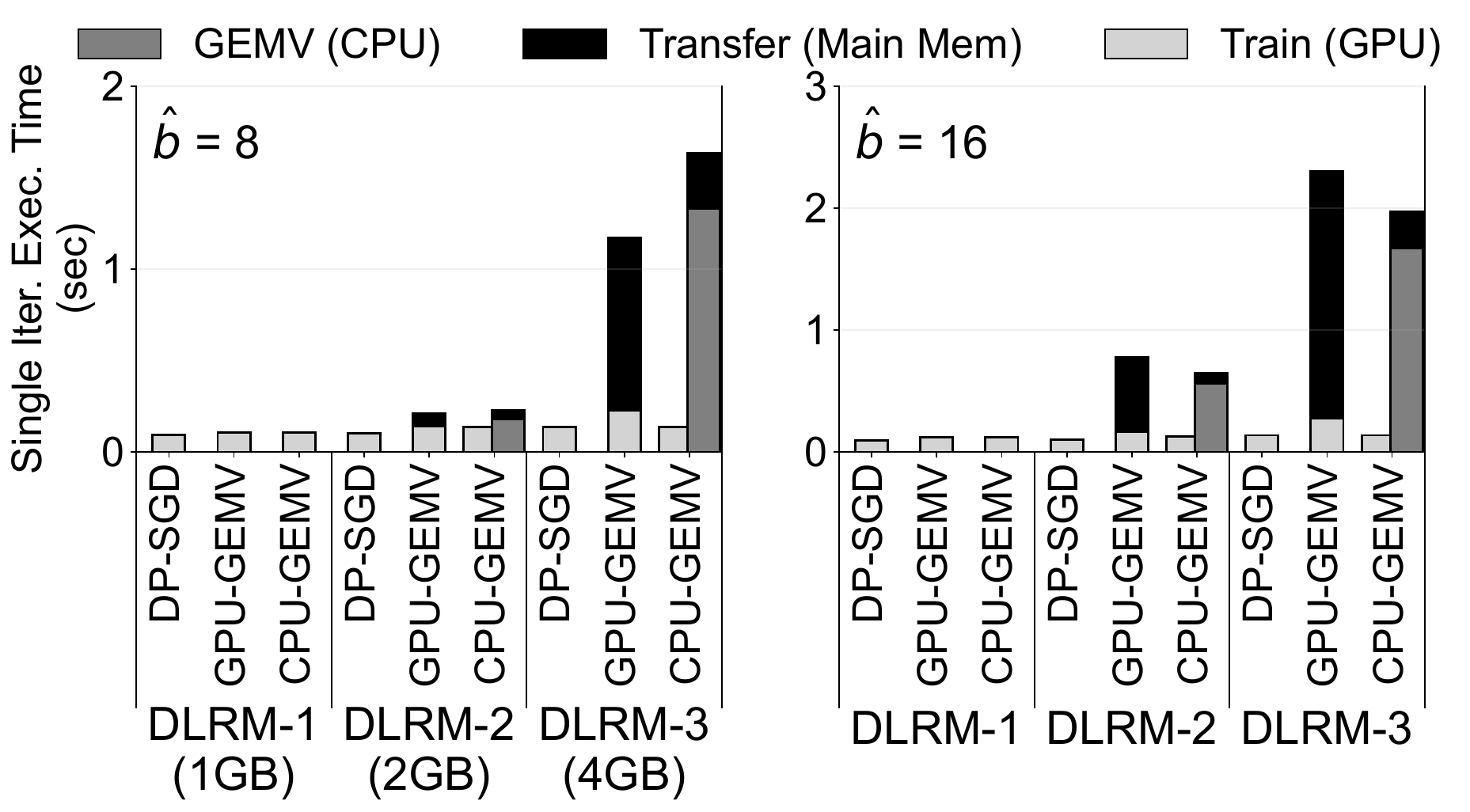}
  \caption{Training time breakdown for DLRM.
  }
  \label{fig:precompute_spoiler}
\end{figure}


\paragraph{Overheads of DLRM.}
Correlated noises incur a unique overhead on DLRMs due to their large embedding tables (Section~\ref{subsubsec:BackgroundDLRM}).
%
Figure~\ref{fig:precompute_spoiler} shows the training time of three DLRMs with different embedding dimensions (DLRM-1/2/3). 
We only characterized a single-GPU setup because the single-iteration latency was too small (near 100ms) to make data parallel training effective.
\otfgpu's latency consists of the GPU-side training and GEMV (``Train (GPU)''), and data transfer from the main memory (``Transfer (Main Mem)''). 
\otfcpu's latency is governed by the slower of the two parallel tasks shown in side-by-side bars: the GPU-side training, and the CPU-side GEMV (``GEMV (CPU)'') plus the transfer of the GEMV result to the GPU (``Transfer (Main Mem)'').
%


In general, \otfgpu is better when the model and/or $\hat{b}$ is small, and \otfcpu outperforms when they are larger. 
However, except for uninteresting cases where the entire noise history fits into GPU (DLRM-1), both baselines still incurred non-negligible slowdown.
%
Even the better baseline between the two incurred \textbf{2.03--8.62$\times$} ($\hat{b}$=8) and \textbf{6.28--14.49$\times$} ($\hat{b}$=16) slowdown.
This was due to the noise-related overheads growing much more linearly with $m$ compared to the training time, which grows sub-linearly with $m$ (Section~\ref{subsubsec:BackgroundDLRM}).
%
%

\emph{
Takeaway 3: For DLRM, correlated noises incur significant slowdown compared to using independent noises.
}



\begin{figure}[t]
  \centering
  \includegraphics[width=0.45\textwidth]{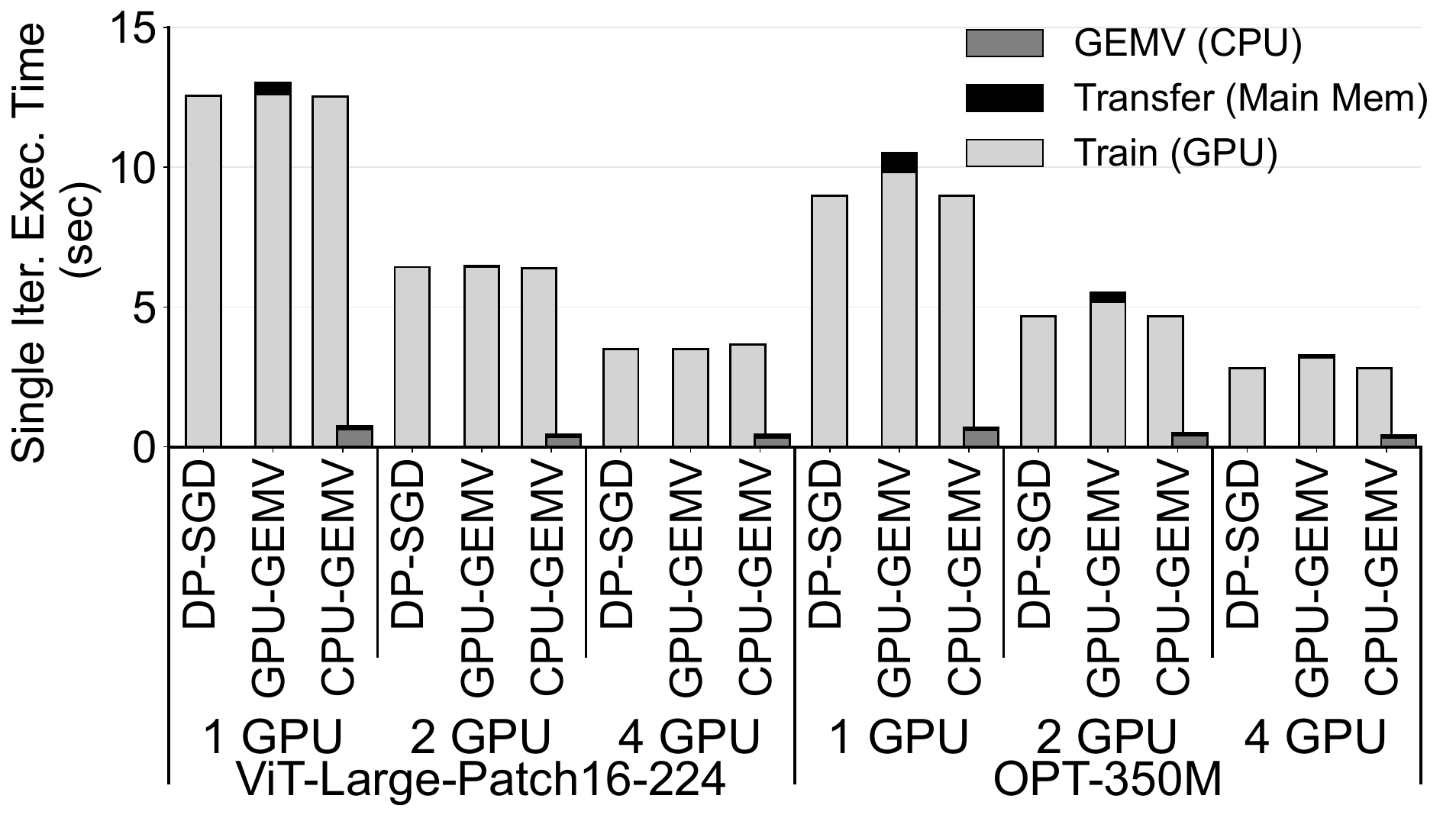}
  \caption{Training time breakdown when the noise history entirely fits into main memory.}
  \label{fig:llm_characterize}
\end{figure}

\paragraph{Overheads of non-DLRM.}
Models other than DLRM experienced similar trends: the exact type mattered less, and their absolute size ($m$) mattered more.
Specifically, the behavior depended on whether the noise history overflowed the main memory and involved secondary memory.
%
%
%
%


Figure~\ref{fig:llm_characterize} shows the result from training ViT-L and OPT-350M with $\hat{b}$=16. In these setups, the entire noise fits into the main memory. The slowdown was only 0.6--18.2\% for \otfgpu, and there was negligible slowdown for \otfcpu.
Other small models (CNNs, ViTs, and sub-billion LLMs) showed similar behaviors, which we omit.
%
%
%



\begin{figure}[t]
    \centering
    \includegraphics[width=0.45\textwidth]{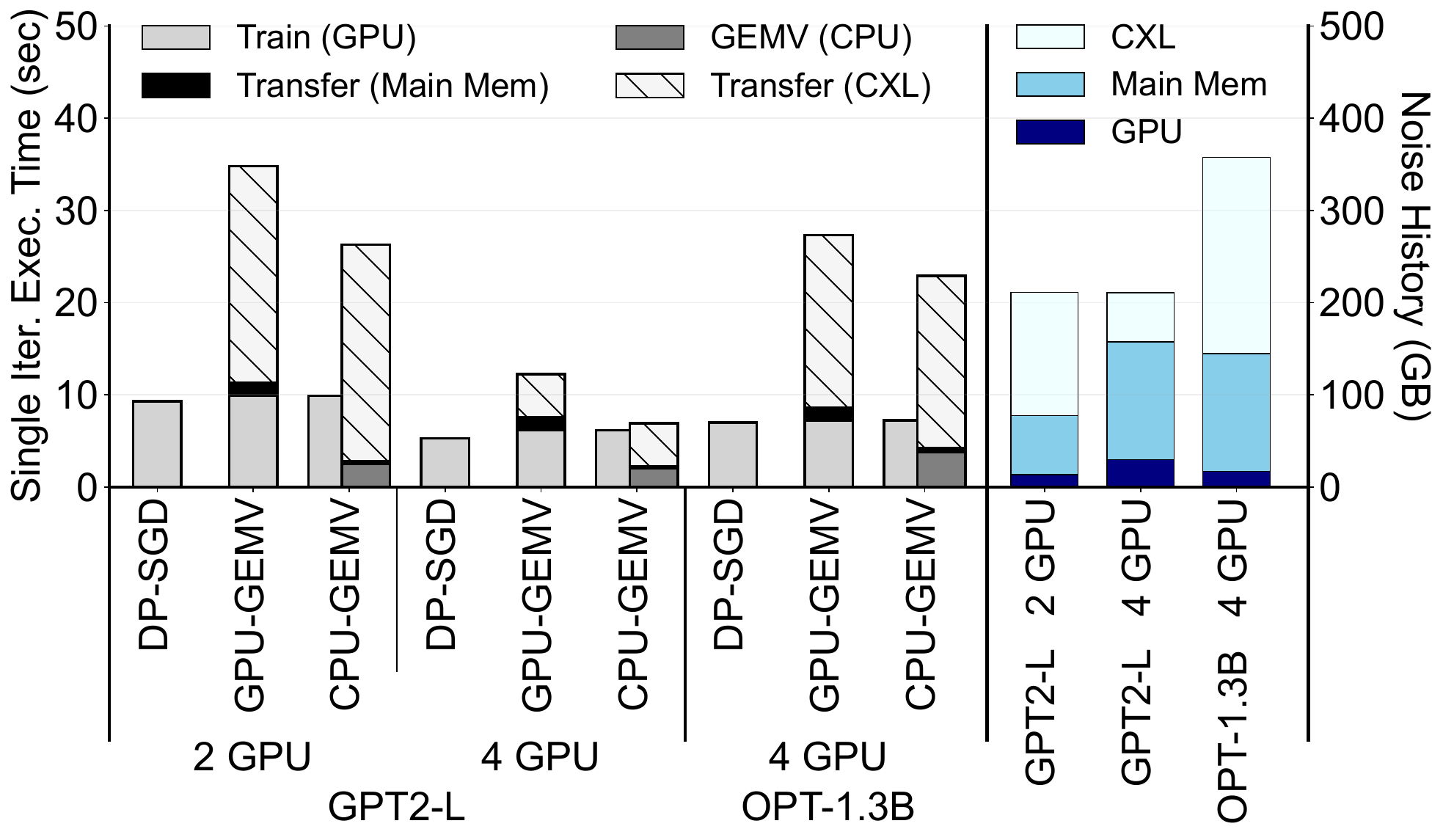}
    \caption{Training time breakdown with noise history partially stored in CXL memory.
    }
    \label{fig:NMPMotivation}
\end{figure}

When the model size and $\hat{b}$ get larger, noise history must be partially offloaded to secondary memory which incurs a non-negligible slowdown. 
%
We only focused on CXL memory due to its relatively better read speed. Slowdown will be higher for slower alternatives like SSD.
Figure~\ref{fig:NMPMotivation} shows the training time breakdown (left), along with the information on where different portions of the noise history are stored (right).
%
%
When training GPT2-L with 2 GPUs, 63\% of the noise history was placed in CXL memory, leading to \textbf{2.83--3.7$5\times$} slowdown.
%
%
When more GPUs are added (GPT2-L with 4 GPUs), some noises could move from CXL memory to GPU memory, improving the slowdown  to 1.30--2.31$\times$.
With a larger OPT-1.3B, the slowdown again increases to \textbf{3.28--3.91$\times$}.
%

\emph{Takeaway 4: 
For non-DLRM models, the overhead is small if the noise history can fit into main memory. When the noise history overflows main memory, data traffic to secondary memory adds significant latency.
}

\paragraph{When CPU is highly utilized.}
We additionally note that \otfcpu may incur a larger slowdown when the CPU suffers from resource contention.
%
As an illustration, we trained OPT-350M with $\hat{b}=64$ on 4 GPUs, and varied the number of cores used by the CPU-side GEMV.
%
We observed that the training starts to slow down if not enough CPU cores can be dedicated to GEMV.
With only 4--7\% of the cores, the training slowed down by 1.52--2.77$\times$.
%



\emph{Takeaway 5: \otfcpu may incur additional overhead when the CPU is highly congested. 
}

%% file: Chapters/4_Coalescing.tex
\section{The \sys System}
\label{sec:Cocoon}
\begin{figure}[t]
    \centering
    \includegraphics[width=0.48\textwidth]{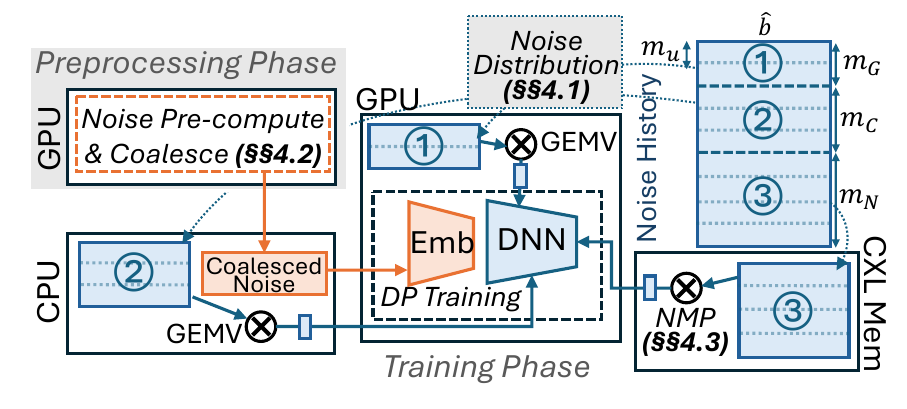}
    \caption{Overview of \sys. 
    \sys distributes the large noise history across \textcircled{1} GPU, \textcircled{2} CPU, and \textcircled{3} CXL memory, and performs GEMV near each memory unit to minimize data movement (Section~\ref{sec:overview}). \sys introduces a preprocessing phase for large embedding tables (Section~\ref{sec:Cocoon-DLRM}), and GEMV inside the CXL memory is done through NMP (Section~\ref{sec:Cocoon-nmp}).
    }
    \label{fig:cocoon_overview}
\end{figure}

We introduce \sys, a framework for efficient DP training with correlated noise.
\sys stores and processes large noise history across GPU, CPU, and CXL memory device with NMP to maximize performance on an economical setup (Figure~\ref{fig:cocoon_overview}).
For large embedding tables of DLRM, \sys provides a dedicated optimization that pre-computes and stores correlated noises in a coalesced format (Section~\ref{sec:Cocoon-DLRM}).
When the CXL memory with NMP exists, \sys offloads part of the GEMV to the NMP device to further improve performance (Section~\ref{sec:Cocoon-nmp}). 
\sys is built on top of Amazon's fastDP~\cite{bu_2023_fastdp_bk}
and tested on a node with multiple GPUs and an industry prototype CXL NMP device.
\sys can leverage other NMP devices as long as they support GEMV~\cite{samsung_nmp, cmm-ax}.

\paragraph{Threat model.} 
DP training methods provide guarantees against an adversary who can access (1) the final model and (2) all the intermediate gradients.
\sys assumes the same adversary except for its embedding table optimization (Section~\ref{sec:Cocoon-DLRM}), where it assumes a weaker-but-practical adversary who can access only the final model and not the intermediate gradients.
%
%
%
%
Assuming an attacker with full knowledge about the intermediate gradients is an artifact of DP-SGD's underlying math and is
often considered excessively strong~\cite{nasr_2021_adversary}. 
The weaker-but-practical adversary we assume for the embedding table optimization reflects the most common real-world threat, such as attackers attempting to extract training data from ML services via APIs~\cite{nasr_2025_chatgpt_attack} or through open-sourced weights~\cite{carlini_2021_gpt2_attack, carlini_2023_diffusion_attack}, where intermediate gradients are unavailable. Many other works adopt this same practical adversary model~\cite{Ning_2022_eana, lim_2024_lazydp, nasr_2021_adversary, dp_iteration, milad_attack, label_only_attack, chuan_attack}.

\subsection{Distributing Noise with Profiling}
\label{sec:overview}
\sys splits the $\hat{b}\times m$ noise history in its parameter ($m$) dimension and distributes slices of size $\hat{b}\times m_G$, $\hat{b}\times m_C$, and $\hat{b}\times m_N$ to GPU, CPU, and NMP device, respectively ($m_G+m_C+m_N=m$).
The goal is to minimize the overall training latency while ensuring none of the devices experience an out-of-memory error.
\sys achieves this through simple profiling, whose cost can be amortized across multiple training iterations.

The goal is solve the following optimization problem:
\[
\begin{aligned}
\min_{m_G, m_C, m_N} \quad & \max\biggl(\lattraingpu + \latgpu(m_G, \hat{b}), \\
& \qquad\quad \latcpu(m_C, \hat{b}), \latnmp(m_N,\hat{b})\biggr) \\
\text{s.t.} \quad & m_G+m_C+m_N=m, \\
                 & \memtotalgpu \ge \memtraingpu + \size(m_G\times \hat{b}) + Buf, \\
                 & \memtotalcpu \ge \memtraincpu + \size(m_C\times \hat{b}), \\
                 & \memtotalnmp \ge \size(m_N\times \hat{b}),
\end{aligned}
\]
\\
where $\lattraingpu$ refers to the single training iteration latency on the GPU, $\latX(m_X, \hat{b}), X\in\{G, C, N\}$ is the GEMV latency on GPU (G), CPU (C), and NMP (N) with the given noise history, $\memtotalX, X\in\{G, C, N\}$ refers to each device's total memory capacity, and $\memtrainX, X\in\{G, C\}$ is the peak memory usage of GPU and CPU for training (NMP is not used for training). 
$Buf$ is a space reserved in GPU for receiving the GEMV result from CPU and NMP plus avoiding out-of-memory (OOM) due to noise in the estimation, and is a design parameter that depends on the implementation.
In short, \sys minimizes the overall latency, which is bottlenecked by the slowest among the three devices (GPU training + GEMV, CPU GEMV, and NMP GEMV), while ensuring no out-of-memory.

Finding an optimal split requires accurately estimating each term. $\size(m_X\times \hat{b})$ and $\memtotalX$ can be known exactly offline, and $\lattraingpu$ and $\memtrainX$ can be upper-bounded by running a few training iterations.
$\latX$ cannot be easily estimated because the GEMV latency is not necessarily linear with the matrix size, especially when the matrix is small. 
%
We simplify the optimization goal by defining a minimum unit of allocation, $\munit$, that perfectly divides $m$, and choose $m_G$, $m_C$, and $m_N$ to be a multiple of $\munit$.
This simplifies the GEMV latency estimation because if $\munit$ is large enough, GEMV becomes memory-bottlenecked and almost linearly scales with the number of $\munit$ processed in each device, \emph{i.e.}, $\latX(m_X, \hat{b}) \approx \frac{m_X}{\munit}\times\latX(\munit, \hat{b})$.
Measuring $\latX(\munit, \hat{b})$ for $X \in \{G, C, N\}$ once and plugging the approximation back in the optimization objective gives us a linear objective that can be solved to get $m_G$, $m_C$, and $m_N$.
Our implementation also receives GEMV results from CPU and NMP devices to the GPU in $\munit$-granularity, which means that $Buf$ should be at least $2\munit$ to temporarily hold GEMV results sent from CPU/NMP.

%
%
%
%
%
As an example, consider training OPT-1.3B with $\hat{b}$=64 on 4$\times$A5000 GPUs (24GB each, $\memtotalgpu$=96GB across 4 GPUs), 128GB DRAM ($\memtotalcpu$=128GB), and a 256GB NMP device ($\memtotalnmp$=256GB). 
The noise of the $m\approx$1.4B model is sharded across the GPUs, CPU, and NMP memory.
Training takes about 72GB across 4 GPUs ($\memtraingpu$=72GB) and 22GB of CPU DRAM ($\memtraincpu$=22GB).
If we heuristically set $m_G$=$m_u$ and $Buf$=1.5GB per GPU, solving the above optimization problem gives $m_u$=$m_G\approx$71M, meaning that $m_u$ divides the parameter $m$ into 20 equal slices. 
GPUs hold one slice across 4 GPUs, CPU holds 5, and NMP can hold the remaining 14.
Varying $Buf$ had little effect on the overall performance, as the GPU-side overhead is mostly due to training, which is independent of $Buf$. Thus, a reasonable value that would not cause OOM suffices.

\subsection{Optimizations for Embedding Tables}
\label{sec:Cocoon-DLRM}

\begin{figure}[t]
    \centering
    \includegraphics[width=0.48\textwidth]{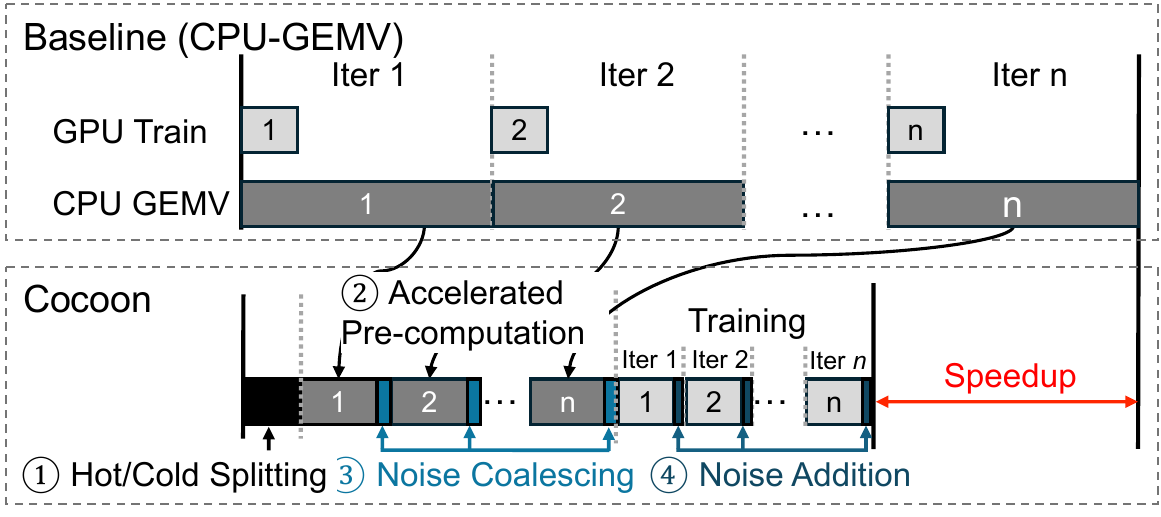}
    \caption{\sys's embedding table optimization. Cocoon replaces the per-iteration CPU-GEMV bottleneck (top) with accelerated noise pre-computation (bottom).}
    \label{fig:cocoon_dlrm_overview}
\end{figure}

%
%
%
Figure~\ref{fig:cocoon_dlrm_overview} summarizes \sys's optimization for embedding tables. For embedding tables, \sys \textcircled{1} splits the table entries by access frequency into either hot or cold, \textcircled{2} efficiently pre-computes all the correlated noises to be used for the cold entries, \textcircled{3} coalesces and stores the noises in a compact format, and \textcircled{4} runs training using the pre-computed noises.

\paragraph{Noise pre-computing.}
Instead of performing GEMV on each iteration (Figure~\ref{fig:cocoon_dlrm_overview}, top), \sys \emph{pre-computes} correlated noises for all the future iterations of the embedding tables before the actual training (Figure~\ref{fig:cocoon_dlrm_overview}, bottom).
%
The pre-computed results cannot be reused across jobs for privacy, so pre-computing must be done efficiently to not bottleneck the entire training.
While pre-computing performs the same amount of GEMV as the baselines, it can be done much faster due to two benefits. Compared to \otfcpu, pre-computing can use the faster GPU, which idles before the training starts.
Compared to \otfgpu, \sys maximizes data reuse inside the GPU and minimizes data transfer through PCIe with \emph{noise tiling}.

\begin{figure}[t]
    \centering
    \includegraphics[width=0.45\textwidth]{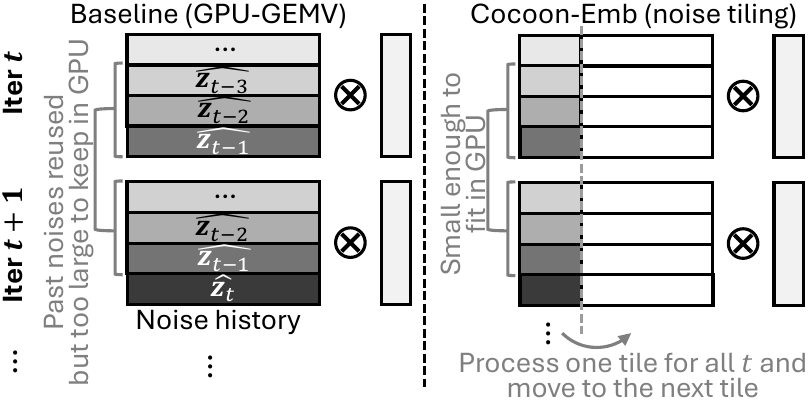}
    \caption{\sys's noise tiling for embedding tables. Cocoon (right) tiles the noise history to fit within GPU memory during pre-computation, avoiding the memory spilling required by the baseline (left) when data exceeds GPU capacity.
    }
    \label{fig:tiling}
\end{figure}

%
%

%

Figure~\ref{fig:tiling} explains noise tiling. 
%
Between consecutive iterations, the most recently updated $\hat{b}-2$ out of $\hat{b}-1$ rows of the noise history are reused (Figure~\ref{fig:tiling}, left).
However, the reused data ($(\hat{b}-2)\times m$) is often too large to be kept inside the GPU, and \otfgpu must spill it to main memory between iterations.
Instead, \sys splits the noise history into smaller tiles and performs noise pre-computing for each tile, while choosing the tile size so that the reused data always fits inside the GPU (Figure~\ref{fig:tiling}, right).
After generating all future noises from one tile, \sys moves on to the next tile.
Noise tiling is only possible during pre-computing, where we can choose to compute noises for all future iterations for one tile before computing the next tile. \otfgpu cannot benefit from noise tiling, as it must immediately compute the entire noise 
(\emph{i.e.}, for all the tiles) 
before proceeding to the next iteration.

\paragraph{Noise coalescing.}
%
%
Without any optimization, the size of the entire pre-computed noises to be used over $n$ training iterations is $m \times n$, which is too large to store.
\sys uses a technique called \emph{noise coalescing} to solve this issue.
As discussed in Section~\ref{subsubsec:BackgroundDLRM}, only a few entries in embedding tables are used at each iteration, and unused entries do not contribute to that iteration's gradient.
Thus, we do not need to accurately update (\emph{i.e.}, add proper noise) all the entries in every iteration. Instead, it is sufficient to add an equivalent, \emph{aggregated} or \emph{coalesced} noises, as long as they are added before an entry is accessed.
%
%

 \begin{figure}[t]
     \centering
     \includegraphics[width=0.42\textwidth]{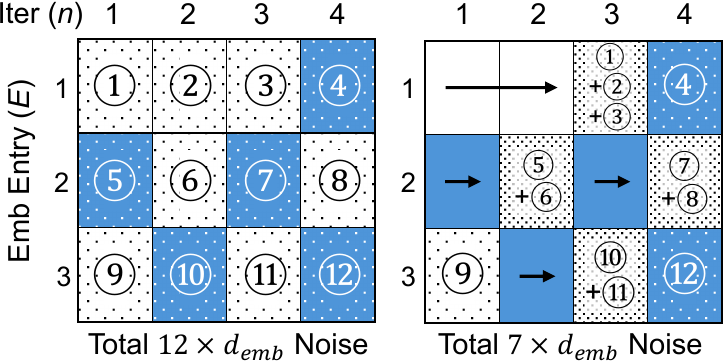}
    \caption{Noise coalescing ($d_{emb}$: embedding dimension). Cocoon (right) reduces storage overhead (12 to 7 units) by aggregating noise until an entry is accessed (filled blue), replacing the per-iteration baseline (left).}
     \label{fig:noise_coalescing}
 \end{figure}

Figure~\ref{fig:noise_coalescing} shows a toy example of an embedding table with three entries trained over four iterations.
Colored boxes indicate in which iteration each entry is accessed, and dotted numbered boxes indicate when noises are added to each entry's gradient.
%
For example, entry 1 is only accessed in the 4th iteration.
Without coalescing (Figure~\ref{fig:noise_coalescing}, left),  noises must be added to all the entries in all the iterations. 
%
Instead, noise coalescing (Figure~\ref{fig:noise_coalescing}, right) only adds an equivalent, aggregated noise right before each entry is accessed or training ends.
%
%
For example, no noise is added to entry 1 in iterations 1--2, and an equivalent noise (\textcircled{1}+\textcircled{2}+\textcircled{3}) is added at the end of iteration 3.
%
%
During pre-computing, \sys merges and only stores the aggregated noise (\emph{e.g.}, stores \textcircled{1}+\textcircled{2}+\textcircled{3} instead of storing three noises separately). In our toy example, only 7 (instead of 12) aggregated noises need to be stored. The benefit is much larger for real models.

%
%

Implementing noise coalescing requires knowing exactly when each entry will be accessed during training. This can be known by using a random batch sampler with the same random seed both during pre-computing and training.
\sys stores the coalesced noise (Figure~\ref{fig:noise_coalescing}, right) in a compressed sparse column (CSC) format.
\sys does not pre-compute noises for the rest of the model (\emph{e.g.}, MLP layers) and simply uses \otfgpu/\otfcpu as they are small.

%

\begin{figure}[t]
    \centering
    \includegraphics[width=0.45\textwidth]{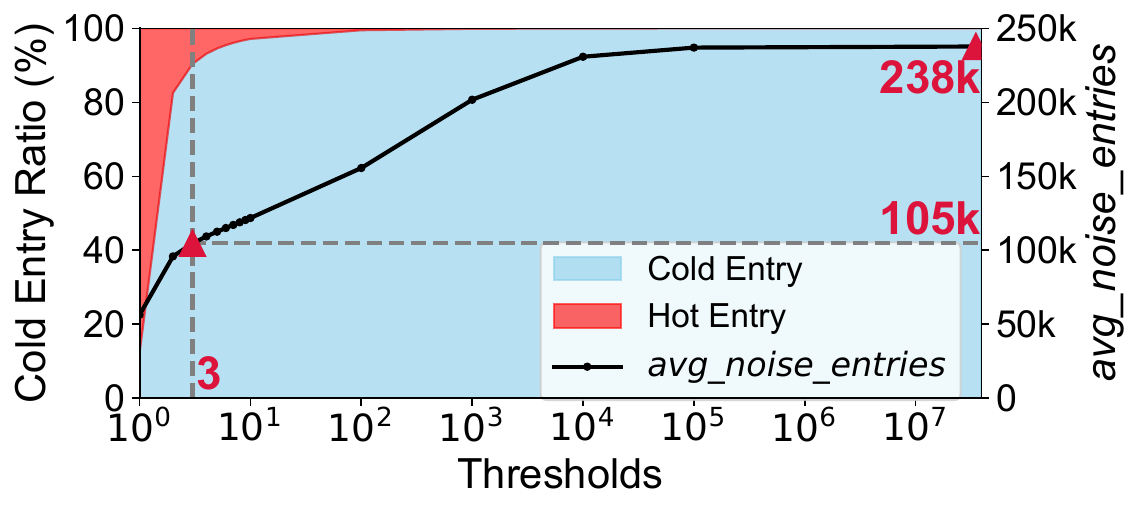}
    \caption{Hot/Cold threshold analysis. 
    Lowering the threshold to 3 reduces \textit{avg\_noise\_entries} (memory overhead) by 2.3$\times$ compared to not using hot/cold splitting.
    }
    \label{fig:HotColdRatio}
\end{figure}
\paragraph{Hot/cold splitting.}
The size of the coalesced noise is directly proportional to 
$\textit{avg\_noise\_entries}$, the average number of entries that need noise to be added in each iteration
($\textit{avg\_noise\_entries}$=$\frac{7}{4}$ in our example),
%
%
%
which is correlated with the average access frequency of each entry.
Typically, most entries are scarcely accessed, but a few ``hot'' entries~\cite{maxim_2019_dlrm} drive up $\textit{avg\_noise\_entries}$.
To reduce $\textit{avg\_noise\_entries}$, \sys classifies each entry as either ``hot'' or ``cold'' and only pre-computes and coalesces noise for cold entries.
Hot entries, just like MLP layers, rely on \otfcpu/\otfgpu.
%
As there are usually only a few hot entries~\cite{maxim_2019_dlrm}, the additional overhead is moderate.
%
We use a simple threshold to classify entries as hot or cold based on their access frequency.
%




Figure~\ref{fig:HotColdRatio} illustrates the relationship between the threshold and the $\textit{avg\_noise\_entries}$ for Criteo Kaggle~\cite{kaggle-2014} dataset.
The dataset has 39 million samples, and the model used for this dataset has 33 million unique embedding table entries.
A lower threshold labels more entries as hot.
%
For example, using 3 as a threshold labels 7\% of the entries as hot, lowering $\textit{avg\_noise\_entries}$ from 238K to 105K (2.3$\times$ memory reduction), compared to not using hot/cold splitting.
We empirically choose the threshold to balance the memory overhead and additional GEMV overhead.

%% file: Chapters/5_NMP.tex
\subsection{Leveraging Near-Memory Processing}
\label{sec:Cocoon-nmp}

When the noise history is too large and must be partially stored in CXL memory, \sys leverages a near-memory processing (NMP) device if available to greatly reduce the data movement overhead. In particular, we demonstrate the benefit by adopting an FPGA-based industry prototype CXL memory device with NMP capabilities, provided through our industry partner.
%
The benefit of adopting an NMP device will become larger for cases where the CPU is heavily utilized.
%

%


\begin{figure}[t]
    \centering
    \includegraphics[width=0.4\textwidth]{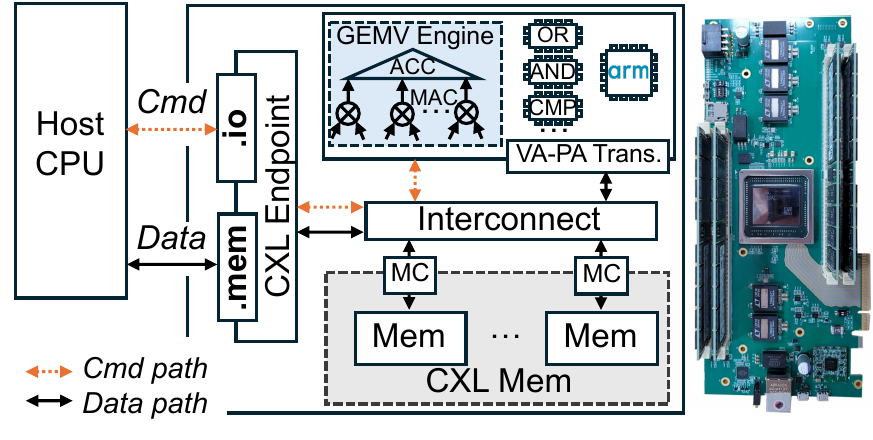}
    \caption{NMP prototype. FPGA-based CXL board featuring an integrated GEMV engine for near-memory acceleration.}
    \label{fig:NMP_HW}
\end{figure}


%


%
\paragraph{NMP device overview.}
Figure~\ref{fig:NMP_HW} shows the prototype device from our industry partner.
The prototype is implemented as an add-in card (AIC)-type custom board that integrates a CXL controller and an NMP engine into a Xilinx Versal (VP1502) FPGA (Figure~\ref{fig:NMP_HW}, right). The board is equipped with DDR4 mounted in DIMM slots. 
The device receives commands from the CPU via CXL.io and data via CXL.mem, and can act either as normal CXL memory or perform GEMV with data in its memory.
The hardware is equipped with MAC and ACC (accumulation) hardware IP, which \sys uses to run the desired GEMV.
%
The prototype additionally contains the complete set of basic logic blocks required to compose SQL and ML operators, which \sys did not use. 
The hardware performs a simple virtual-to-physical address translation through 
saving and looking up memory offset for each matrix and using it to locate them in CXL memory.
%
%
%
%
When there are multiple jobs running on the host, their commands are queued and processed in a first-come-first-served fashion.

\begin{figure}[t]
    \centering
    \includegraphics[width=0.4\textwidth]{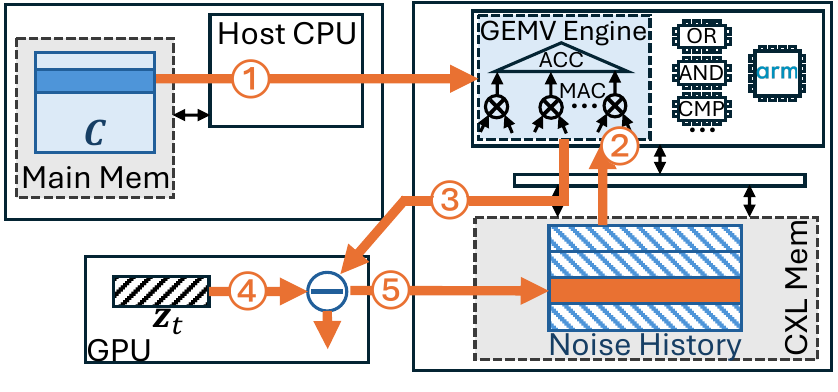}
    \caption{Correlated noise generation with NMP. CPU offloads GEMV to NMP (\textcircled{1}-\textcircled{3}) with Gaussian noise generation (\textcircled{4}) to produce correlated noise that updates the noise history and model (\textcircled{5}).}
    \label{fig:nmp_flow}
\end{figure}

\paragraph{Integration to \sys.}
Figure~\ref{fig:nmp_flow} illustrates how correlated noise generation is partially offloaded to the NMP device.
The $\hat{b}-1$ past noises have the size of a $(\hat{b}-1)\times m$ noise history matrix. 
Noise used at step $t$ is stored at the
$(t\Mod{(\hat{b}-1)})$-th row, updating the rows in a circular manner (\emph{i.e.,} storing noise history in a ring buffer). 
At each iteration, the CPU \textcircled{1} passes an appropriate mixing vector to the NMP device, \textcircled{2} initiates GEMV between the mixing vector and the noise history, and \textcircled{3} the GEMV result sent to GPU. \textcircled{4} the GPU generates Gaussian noise (noise generation can be done by others as well for perfect parallelism, but we use the GPU due to its high throughput).
Finally, \textcircled{5} the generated noise updates the noise history and model.
These process can be done in parallel while the GPU performs training.
\sys pre-normalizes the mixing vector ($\mathbf{C}[t, t-\tau]$ in Equation~\ref{eq:bandedmf}) and the Gaussian noise ($\mathbf{z}_t$) by the $(t,t)$-th entry of $\mathbf{C}$ prior to GEMV to avoid later scaling.
As the noise history table is updated in a circular fashion, the mixing vector must also be properly reordered, which is done statically before training.

\sys sends the mixing vector once to the NMP device, stores it in an internal buffer, and reuses it $m$ times. For reasonably large models, this amortizes the vector transfer cost.
The GEMV operation is implemented with MAC and ACC (accumulation) hardware IP, and the memory bandwidth is maximized through memory-channel interleaving. 
The prototype device we used achieved 47.9GB/s peak GEMV throughput. The performance may improve in the future with faster DRAM technologies (\emph{e.g.}, DDR5).
While the current implementation detail is specific to the prototype we used, we believe similar integration will be possible with other NMP devices~\cite{ko2025cosmos, park_2024_lpddr, sim2022computational}, as long as they allow GEMV. 
Coherence is controlled by flushing at the end of every iteration.

%% file: Chapters/7_evaluation.tex
\section{Evaluation}
\label{sec:Evaluation}


\subsection{Experimental Setup}
\label{sec:experimental_setup}

\noindent \textbf{Hardware.}
Most of our evaluation was done on a dual-socket Intel Xeon Gold 6330 CPU with 256GB DRAM and 8 NVIDIA RTX A5000 GPUs. 
When needed, we ran additional experiments on a more powerful dual-socket AMD EPYC 7763 CPU server with 1TB of DRAM and 8 NVIDIA A100 (80GB) GPUs. 
%
We additionally used the industry prototype NMP device.
Unless noted otherwise, DLRMs were trained on a single GPU without NMP, and LLMs were trained on four GPUs with NMP.
CPU-side GEMV used Intel MKL (Intel) and OpenBLAS (AMD) in Pytorch (v.2.4.0).
%

Due to its early-stage engineering, our NMP prototype exhibited a suboptimal memcpy throughput of 5–7GB/s. This is significantly lower than what a typical CXL memory can achieve (32GB/s for gen5 8x).
To \emph{emulate} the end-to-end performance of a mature system, we separately measured the NMP device's overhead and analytically scaled the memcpy overhead assuming a 22GB/s memcpy throughput, which was measured from a similar in-house CXL memory.
During training, we run a separate process that adds the estimated NMP device overhead to the training latency, while the rest of the CPU/GPU runs training end-to-end. 
Although the end-to-end performance is emulated when NMP is involved, except for the memcpy performance, all the overheads were from real measurements.
When NMP is not involved, our results are from direct end-to-end measurement.
%
%
We assume all resources (CPU cores, PCIe bandwidth, CXL memory capacity, \emph{etc.}) are divided evenly across GPUs.


\noindent \textbf{Datasets/models.}
For DLRM, we used the Criteo Kaggle dataset~\cite{kaggle-2014} and the architecture from~\cite{maxim_2019_dlrm}.
We additionally generated synthetic datasets to study the impact of varying the number of embedding entries and data skewness.
%
Synthetic datasets were generated by first ensuring all embedding entries are accessed at least once, and generating the remainder such that the entry accesses follow a Zipfian distribution with a varying $\alpha$. 
For non-DLRMs, we used ImageNet~\cite{cvpr-2009-imagenet}-sized dummy data and E2E dataset~\cite{novikova-etal-2017-e2e}, although the datasets do not impact performance, and architectures from TorchVision/HuggingFace.

\noindent \textbf{Hyperparameters.}
The training batch size $B$ and band size $\hat{b}$ crucially influence the correlated noise overheads. 
We used the following values from the literature: $B=1024$ for vision and language models~\cite{bu_2022_scalable, bu_2023_fastdp_bk}, $B=65536$ for DLRMs~\cite{chua_2024_scalable}, and $\hat{b}$=2--256~\cite{neurips_2023_amplified, mckenna_2025_scaling, ganesh_2025_dp_optimizers}.
The impact of these hyperparameters is additionally evaluated in the sensitivity studies.

%






\begin{figure}[t]
    \centering
    \includegraphics[width=0.48\textwidth]{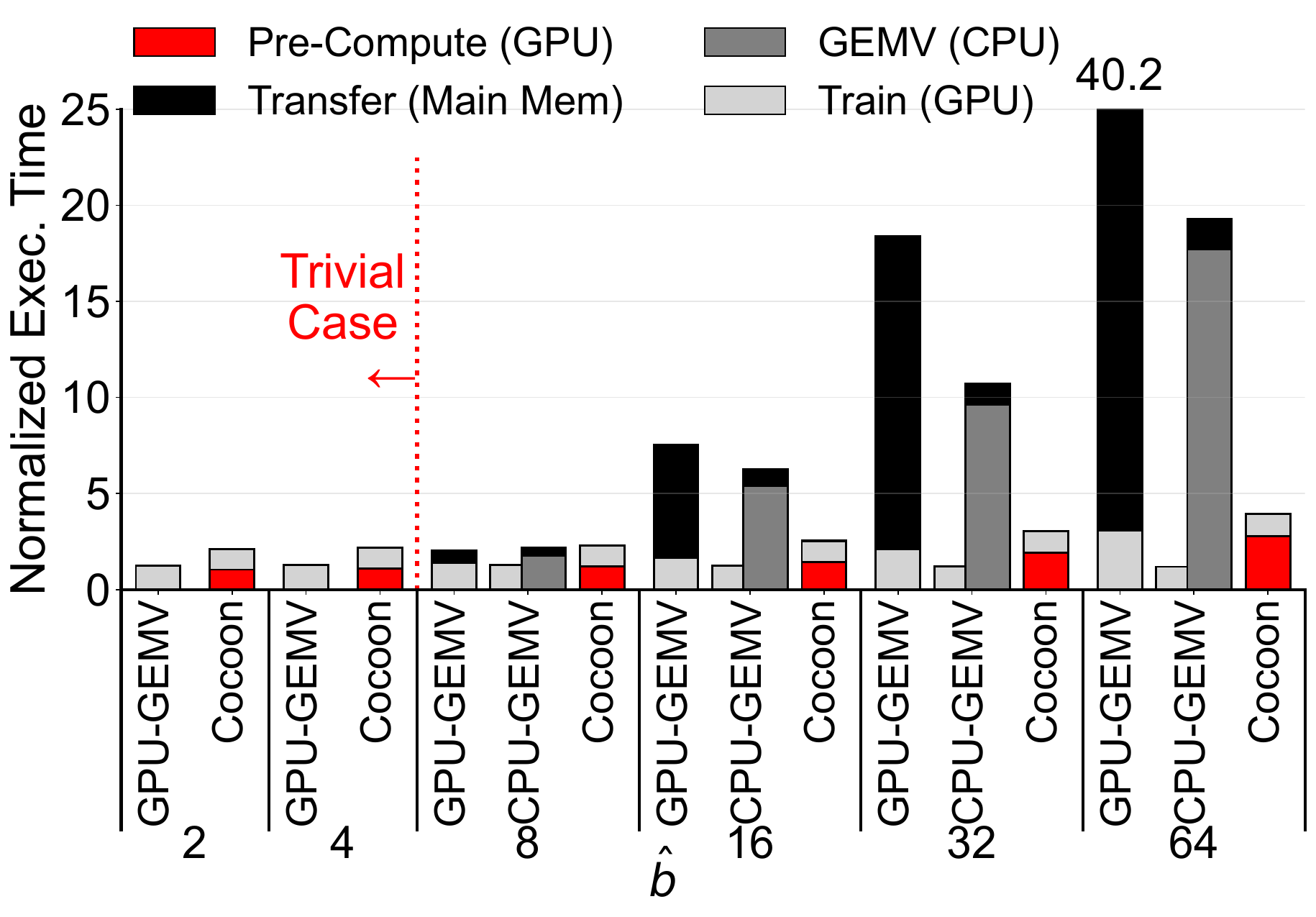}
    \caption{
    Normalized training time of \sys with DLRM.
    When the entire noise history fits into GPU memory (trivial case; $\hat{b}$=2--4), our optimization can be turned off.
    }
    \label{fig:TimeBand}
\end{figure}

\subsection{Performance Improvements: DLRM}
\label{subsec:EvalDLRMTrainingTime}

\subsubsection{End-to-End Training Time}
\label{subsec:E2Eanalysis}

Figure~\ref{fig:TimeBand} compares the DLRM training time of \sys with the baselines. All the bars are normalized to the training time of DP-SGD.
%
%
For $\hat{b} > 8$, \sys consistently outperforms the baselines. Compared to the better baseline, \sys improves the overall training time by \textbf{2.46--4.87$\times$} for $\hat{b} > 8$. The speedup generally increases with $\hat{b}$ (2.46$\times$ for $\hat{b}=16$ and 4.87$\times$ for $\hat{b}=64$).
The breakdown shows that pre-computing dominates the \sysdlrm latency.

When $\hat{b} < 8$, the entire noise history fits into the GPU. 
For these \emph{trivial cases}, the training time of both baselines becomes identical (we only show \otfgpu) and close to DP-SGD, and our optimization simply adds unnecessary overheads.
%
Such trivial cases can be easily detected by comparing the noise history size with the GPU memory capacity, and our optimization can be turned off.
%

\begin{figure}[t]
    \centering
    \begin{subfigure}{0.45\textwidth}
        \centering
        \includegraphics[width=\linewidth]{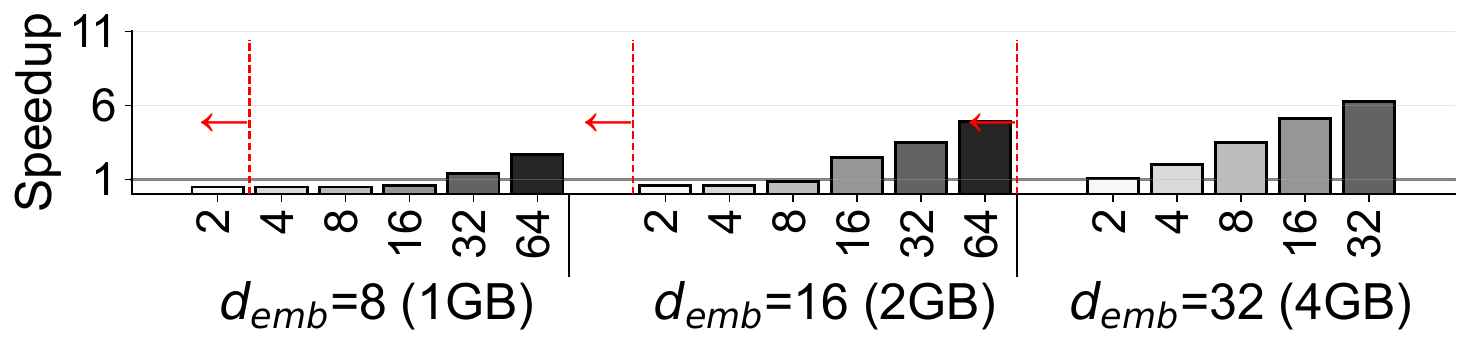}
        \caption{Varying embedding dimension ($d_{emb}$).}
        \label{fig:Sens-Dimsize}
    \end{subfigure}
    \begin{subfigure}{0.45\textwidth}
        \centering
        \includegraphics[width=\linewidth]{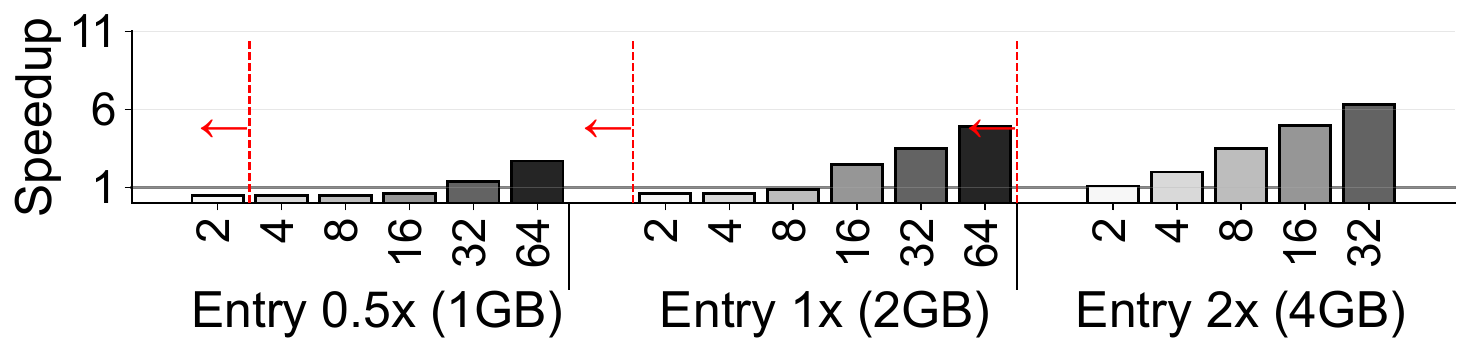}
        \caption{Varying the number of embedding entries.}
        \label{fig:Sens-EntrySize}
    \end{subfigure}
    \begin{subfigure}{0.45\textwidth}
        \centering
        \includegraphics[width=\linewidth]{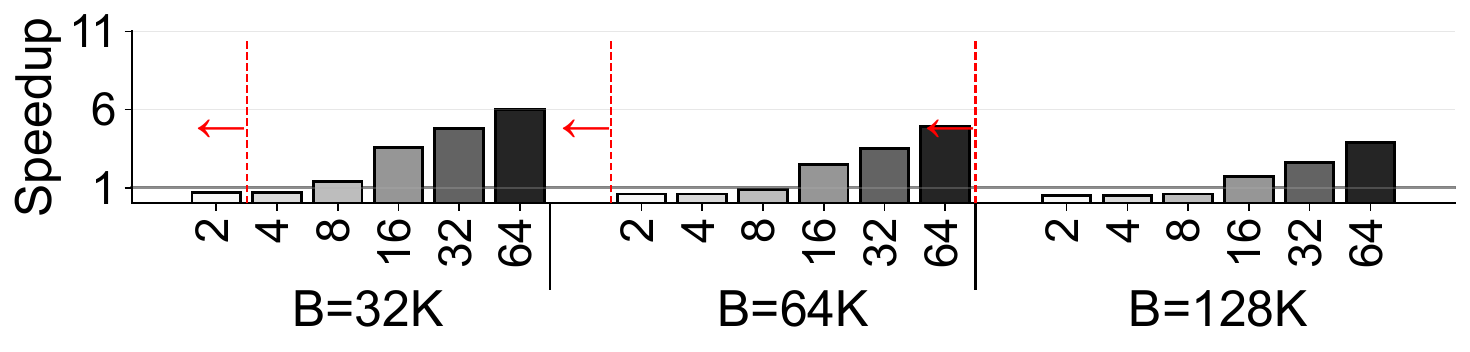}
        \caption{Varying batch size ($B$).}
        \label{fig:Sens-BatchSize}
    \end{subfigure}
    \begin{subfigure}{0.45\textwidth}
        \centering
        \includegraphics[width=\linewidth]{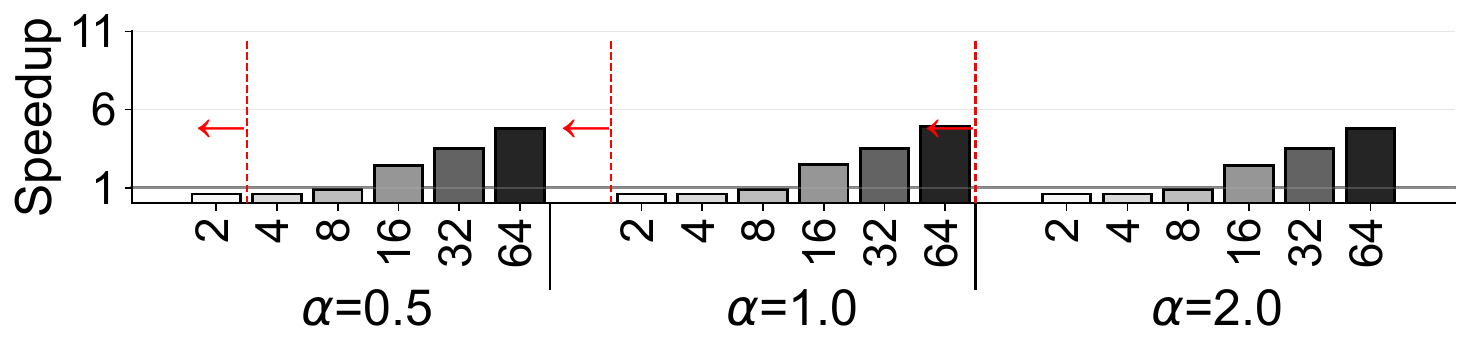}
        \caption{Varying skewness (larger Zipf $\alpha$ means more skewed).}
        \label{fig:Sens-Skewness}
    \end{subfigure}
    
    \caption{Speedup of \sys under various models and datasets.
    Numbers below each bar indicates $\hat{b}$. 
    }
    \label{fig:EvalSensitivity}
\end{figure}

\subsubsection{Sensitivity Study}
\label{subsec:Sensitivity}

Figure~\ref{fig:EvalSensitivity} shows the speedup of \sys over the better baseline between \otfcpu and \otfgpu, while varying different dimensions of the model and dataset. 
Again, bars located at the left side of the red vertical line are trivial cases, and our embedding table optimizations can be turned off.


\noindent \textbf{Model size.} Figures~\ref{fig:Sens-Dimsize} and \ref{fig:Sens-EntrySize} show the speedup of \sys while varying the model size, adjusting the embedding dimension ($d_{emb}$) or the number of embedding entries.
%
The speedup improves with the model size: if we compare the bars at $\hat{b}=32$, the speedup improves from 3.51$\times$ to \textbf{6.27-6.35$\times$} when the size is doubled, and reduced to 1.37$\times$ when halved.
This is because larger models must offload more noise history to main memory and penalize the baselines more severely. 
\sys will become more effective as models grow.


\noindent \textbf{Batch size.} 
Figure~\ref{fig:Sens-BatchSize} shows that the speedup decreases with an increasing batch size. When considering $\hat{b}=32$, the peak speedup increased from 3.51$\times$ with $B=64K$ to \textbf{4.79$\times$} with $B=32K$, and reduced to 2.57$\times$ with $B=128K$.
%
This is because the correlated noise generation overhead (which \sys optimizes) stays the same regardless of the batch size, while the training latency (which \sys cannot optimize) becomes larger with bigger $B$.
While not shown, Doubling the entries accessed per sample (\emph{i.e.}, pooling factor~\cite{maxim_2019_dlrm}) has nearly the same effect as doubling the batch size.



\noindent \textbf{Skewness.} Figure~\ref{fig:Sens-Skewness} shows the speedup of \sys when the access frequency of each embedding entry experiences different skewness. 
The skewness was controlled by varying $\alpha$ of the Zipfian distribution of our synthetic dataset.
Interestingly, the skewness had only a minor effect on training time.
We will later show that skewness is a critical factor in the memory footprint in Section~\ref{subsec:MemConsumption}.



\begin{figure}[t]
    \centering
    \includegraphics[width=0.47\textwidth]{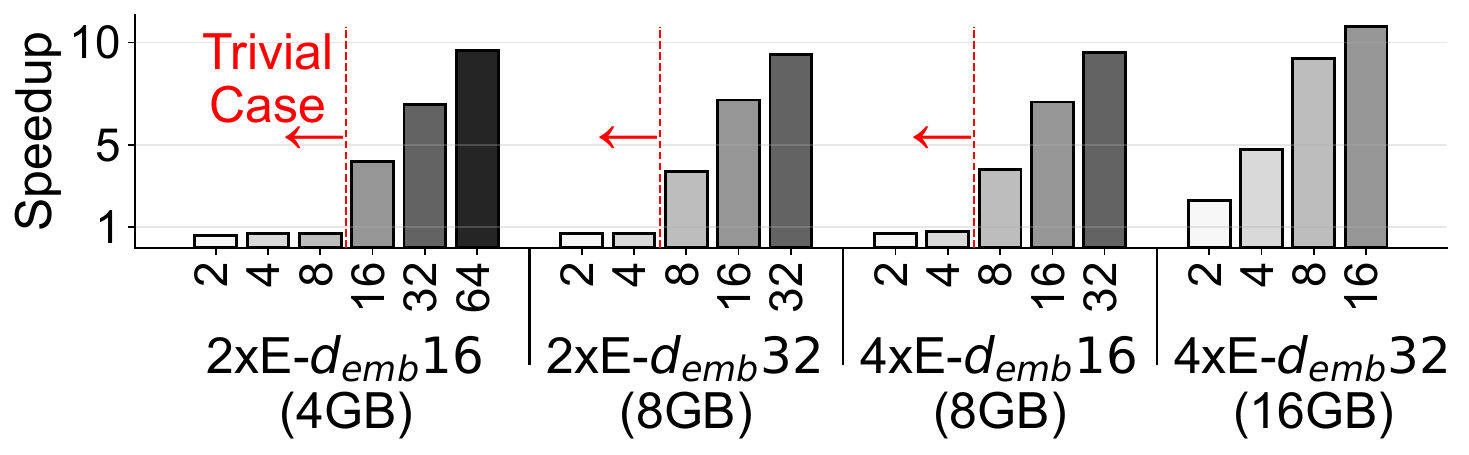}
    \caption{Speedup of \sys with various model sizes on an A100 GPU. 
    Numbers below each bar indicates $\hat{b}$.
    }
    \label{fig:A100Result}
\end{figure}

\noindent \textbf{Hardware.} Figure~\ref{fig:A100Result} shows the speedup of \sys on more powerful A100 GPUs. As A100 has more memory, we used larger models (4--16GB) with a larger number of embedding entries (2--4$\times$ larger, denoted as 2$\times$E/4$\times$E) and bigger embedding dimensions ($d_{emb}$=16--32).
%
\sys achieved a speedup of \textbf{2.33--10.82$\times$} for non-trivial cases, which is generally larger than the results on the A5000.
\sys's speedup is larger on A100 because its computation (GPU-side GEMV and training) is much faster than on the A5000, but overheads related to correlated noises (GPU-main memory data transfer, CPU-side GEMV), which \sys optimizes, are similar.
%

%

\subsection{Memory Overhead: DLRM}
\label{subsec:MemConsumption}

\subsubsection{Overall Memory Overhead}

Unlike \otfcpu and \otfgpu, which incur a fixed $O(\hat{b}m)$ memory overhead, the overhead of \sys with its embedding table optimization depends on the effectiveness of its noise coalescing.
In the worst case, \sys must hold all the pre-computed noises ($O(nm)$ with $n$ iterations), which can be much larger than that of \otfcpu/\otfgpu (usually, $n\! \gg\! \hat{b}$).
However, the actual memory overhead is much less thanks to noise coalescing.
%

Figure~\ref{fig:MemoryConsumption} evaluates the memory footprint of the coalesced noise of \sys while varying the embedding dimension ($d_{emb}$), batch size, number of embedding entries, and entry access distribution skewness.
The bars are normalized to the model size $m$.
%
For this figure, we used $n=1800$ (three epochs using Criteo Kaggle~\cite{kaggle-2014} dataset with $B=64K$), so the worst-case overhead is 1800$\times m$.
%
However, the actual memory overhead is only 4.3--31.6$\times$, which is \emph{less} than the memory overhead of the baselines in many cases (shown in horizontal lines for $\hat{b}$=16 and $\hat{b}$=32). 
%
The memory overhead of \sys is independent of $\hat{b}$ and only depends on the entry access pattern, while the overheads of the baselines grow linearly with $\hat{b}$.
%

\subsubsection{Sensitivity Study}

\begin{figure}[t]
    \centering
    \includegraphics[width=0.45\textwidth]{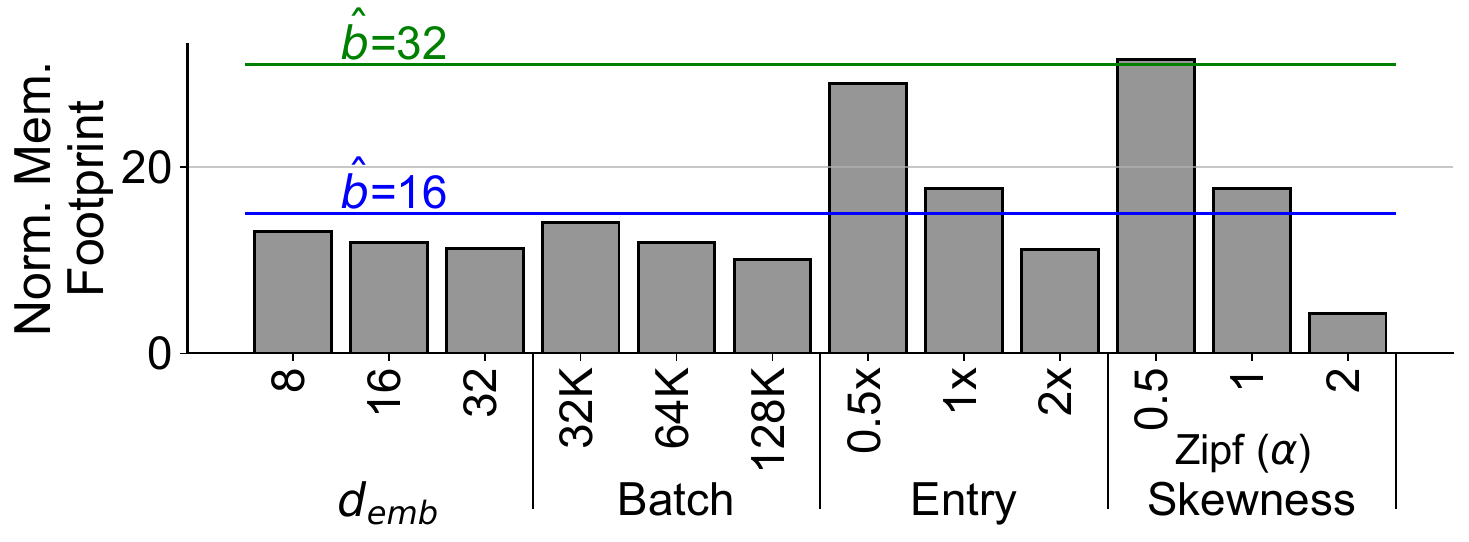}
    \caption{Memory footprint of coalesced noise normalized by the model size. Memory footprint of a noise history without pre-computing for different $\hat{b}$ are in horizontal lines.
    }
    \label{fig:MemoryConsumption}
\end{figure}

Figure~\ref{fig:MemoryConsumption} shows how different models and datasets affect the efficacy of noise coalescing.
It can be seen that the efficacy decreases with reducing $d_{emb}$ and batch size, but the effect is small.
Conversely, decreasing the number of embedding entries and using datasets with less skewed patterns significantly increases the memory overhead.
This meets our expectation because noise coalescing works better when batched samples are mostly accessing the same entries, which leads to lower \textit{avg\_noise\_entries}. 
%
Decreasing the number of entries has a similar effect to reduced skewness, because the accesses are hashed into the remaining entries.

\begin{figure}[t]
    \centering
    \includegraphics[width=0.48\textwidth]{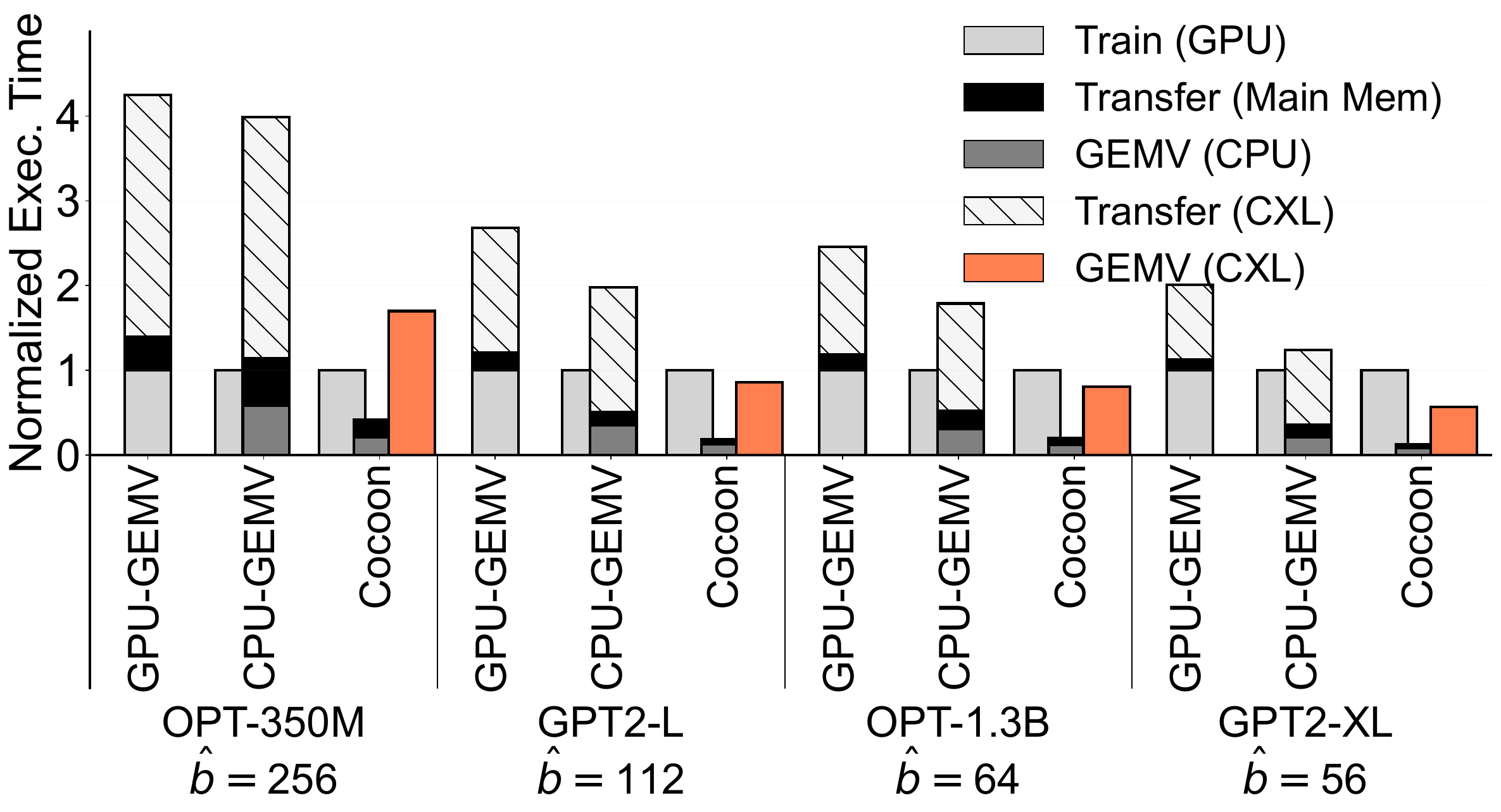}
    \caption{
    End-to-end normalized training time of \sys+NMP and the baselines.
    Model sizes are in an ascending order.
    $\hat{b}$ is chosen, so that over 200GB of the noise history is offloaded to CXL memory. 
    }
    \label{fig:NMP_result_var_models}
\end{figure}

\subsection{Performance Improvement with NMP}

\subsubsection{End-to-End Training Time}
Figure~\ref{fig:NMP_result_var_models} plots the training time and breakdown for the baselines and \sys with the NMP hardware prototype, when the models are large enough to involve CXL memory.
\sys has three bars side-by-side, indicating the GPU-side training, CPU-side GEMV, and the GEMV happening inside the NMP device, all happening in parallel.
$\hat{b}$ values are chosen for each model to ensure over 200GB of the noise history is offloaded to CXL memory, to avoid trivial, uninteresting setups.

Figure~\ref{fig:NMP_result_var_models} shows that \sys consistently outperforms the baselines, achieving \textbf{1.23--2.32$\times$} speedup compared to the better baseline.
\sys achieves high speedup by eliminating the large data transfer overhead between the CXL memory and CPU/GPU (``Transfer (CXL)''), while incurring a moderate GEMV overhead inside the CXL controller (``GEMV (CXL)'').
When the GEMV overhead of \sys is less than the training time and can be completely hidden (GPT2-L, OPT-1.3B, GPT2-XL), training becomes the critical path. 
Our noise distribution strategy minimizes the overall latency, trying to put the minimum noise history on the bottlenecked device.
%
%

\begin{figure}[t]
    \centering
    \includegraphics[width=0.48\textwidth]{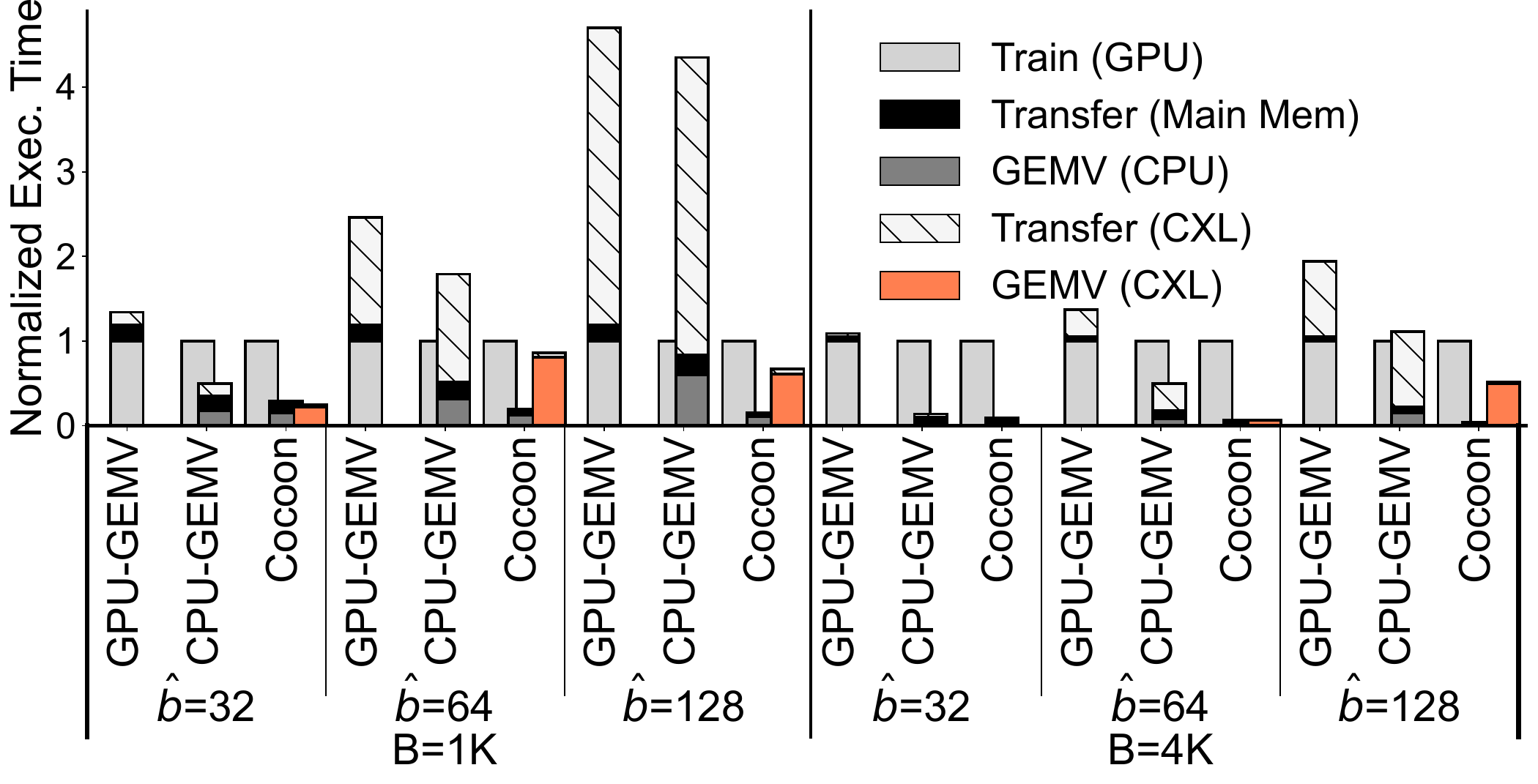}
    \caption{End-to-end normalized training time of \sys+NMP and the baselines with OPT-1.3B.
    $\hat{b}=128$ number assumes a hypothetical CXL device with an expanded memory capacity, as it does not fit into our current CXL prototype.
    }
    \label{fig:NMP_result_opt}
\end{figure}

\begin{table}[]
\small
\centering
\caption{Hardware cost and power estimate (GPT2-XL, $\hat{b}$=64). \label{tbl:cost}}
\begin{tabular}{cccc}
\cline{1-4}
\multicolumn{4}{c}{\begin{tabular}[c]{@{}c@{}} \textbf{1 HW Unit}: 4$\times$NIVIDA A5000 (4$\times$\$2K), \\  1$\times$ Intel Xeon Gold 6330 (1$\times$\$2.1K), \\ 8$\times$16GB DDR4 3200MHz (8$\times$\$0.15K) \end{tabular}} \\ \hline
 & GPU-Only & CPU-GEMV & \textbf{Cocoon} \\ \hline
\begin{tabular}[c]{@{}c@{}}HW Unit\\\end{tabular}& 10  & 3& 1 + NMP~\cite{park_2024_lpddr}  \\ \hline
Cost (\$) & 113K  & 33.9K& 18.3K \\ \hline
\begin{tabular}[c]{@{}c@{}}Peak Power \\ (KW)\end{tabular} &  11.6 & 3.5 & 1.2 \\ \hline
\begin{tabular}[c]{@{}c@{}}Normalized\\ Throughput\end{tabular} & 1 & 0.33  & 0.13 \\ \hline
\end{tabular}
\end{table}

\subsubsection{Power and Cost}
Table~\ref{tbl:cost} compares the estimated hardware cost and peak power of \sys with the NMP device against two alternative setups (GPU-Only, CPU-GEMV), which do not involve CXL memory and instead scale out the GPU (GPU-Only) or CPU (CPU-GEMV) to hold the large noise history.
GPU-Only represents prior works' setups~\cite{mckenna_2024_scaling, pillutla2025correlated} where enough GPUs hold the entire noise history.
We assume GPT2-XL training with $\hat{b}$=64 ($\sim$413GB of noise history).
%
As there can be many different ways to scale out, we simplify the estimation by considering our current single-socket configuration (4xA5000 + 1xIntel Xeon Gold + 128GB memory) as one hardware unit, and estimating how many such units are needed to support noise generation.
%
%
%
%
As our prototype NMP device does not have a public power/cost number, we use numbers from prior work that used a similar device~\cite{park_2024_lpddr}.
The device from~\cite{park_2024_lpddr} is more powerful than our prototype, so the power/cost numbers are likely to be an \emph{upper bound}.
We do not model the extra cost (\emph{e.g.}, network, power supply) of scaling out beyond a single node.
All of these simplistic estimations \emph{favor other baselines}, and \sys's actual benefit is likely to be higher.

Table~\ref{tbl:cost} shows that \sys can realize training with much less power and hardware cost.
This is because instead of using expensive GPU/CPU memory, \sys can use CXL NMP memory to partially host and process noise history, which is expected to be much cheaper and lower power than CPU/GPU~\cite{park_2024_lpddr}.
%
However, as the throughput also increases with the hardware units, the baselines also achieve better throughput.
If you compare the ratio between the normalized throughput numbers and the cost/power, \sys is similar to CPU-GEMV, while being much more cost/power-efficient than GPU-Only. This is because GPU-GEMV introduces too many GPUs, and GPUs can become underutilized.
The result indicates that scaling out CPU/GPU is better when high throughput is the main goal, while \sys with NMP can be useful if one has a limit on the power/cost budget.


\subsubsection{Sensitivity Study}

\noindent \textbf{Model size.}
Figure~\ref{fig:NMP_result_var_models} also shows that smaller models (models to the left) achieve more speedup than larger models (models to the right), when the noise history size inside the CXL memory is similar.
This is because the training time ultimately becomes the major bottleneck and limits achievable speedup in larger models.

\noindent \textbf{Band size.}
Figure~\ref{fig:NMP_result_opt} plots the end-to-end training time and breakdown for OPT-1.3B while varying $\hat{b}$.
When using $\hat{b}=128$, the noise history is too large to fit into our CXL memory,
and we cannot currently run it on our CXL prototype.
For $\hat{b}=128$, we provide an analytical projection assuming a hypothetical future device with expanded memory capacity, but otherwise identical characteristics (GEMV throughput, interconnect bandwidth, \emph{etc.}).
This additional datapoint shows how the idea can scale to larger CXL memory devices in the future to support even larger models and band sizes.
When $\hat{b}$ is small (\emph{e.g.}, $\hat{b}$=32), \sys and \otfcpu both perform similarly to DP-SGD, because the training time dominates and all the other overheads are hidden.
With larger $\hat{b}$, both the baseline and \sys start to incur slowdown, and \sys outperforms the best baseline by 2.26$\times$.

%
%
%


\noindent \textbf{Batch size.} Figure~\ref{fig:NMP_result_opt} shows that the speedup decreases when we increase the batch size from $B$=1K to $B$=4K. 
This is because the training time increases with larger batch sizes, and the rest of the overheads can be hidden behind this increased training time.
With $B$=4K, \sys showed speedup over the best baseline only when $\hat{b}$=64; otherwise, both \sys and the baseline performed on-par.
The result indicates that \sys's benefit will decrease when using larger batches, but would still show speedup when $\hat{b}$ is large enough.

\subsection{Performance Improvement: DLRM+NMP}

Our current evaluation only covers the use of NMP with LLMs, not DLRMs. This is because our NMP prototype's limited GEMV throughput was insufficient to benefit DLRM, whose single-iteration latency was on the order of hundreds of milliseconds. 
Currently, all the GEMV for DLRM (during both pre-compute and actual training) is done on the GPU to operate near this timescale, as detailed in Section~\ref{sec:Cocoon-DLRM}. 
However, DLRM can also benefit from NMP if its GEMV throughput improves in the future. For example, \cite{park_2024_lpddr} claimed that building an NMP device with TB/s of internal bandwidth is possible. 
If such a high bandwidth is indeed possible, we can use the NMP device to accelerate pre-computing of the noise history for the embedding tables. If we assume the throughput from \cite{park_2024_lpddr}, our analytical projection shows that DLRM training would be accelerated by an additional \textbf{2.4$\times$}.

%% file: Chapters/8_related.tex
\section{Related Work}
\label{sec:RelatedWork}

\noindent \textbf{DP-SGD for large models}.
In the earlier days of DP training research, it was thought that DP training only works well for small models~\cite{de_2022_unlocking}.
Recently, many studies have shown that DP training can work well for larger foundation models~\cite{zhang2024disk, dp_llm_2, mckenna_2025_scaling, dp_llm_4, dp_llm_5, dp_llm_6, dp_llm_7, wang2024flashdp, chua_2024_mind, charles_2024_fine_tuning, dp_llm_11, beltran_2024_towards, bu_2023_fastdp_bk} and DLRMs~\cite{ghazi_2023_adafest, dp_recsys_2, dp_recsys_3, chua_2024_scalable, Ning_2022_eana, lim_2024_lazydp} as well.
Google recently released an LLM that was fully trained from scratch with DP~\cite{sinha_2025_vaultgemma}.


\noindent \textbf{System optimizations for DP training.} Several works studied how DP-SGD can be made faster through optimizing the software~\cite{opacus_2020, bu_2023_fastdp_bk, lee_2021_scaling, bu_2022_scalable} and hardware~\cite{park_2022_diva}. These works mainly focused on efficiently calculating the per-example gradient and are orthogonal to this work.
There are also works that studied how to accelerate DP-SGD for DLRMs~\cite{Ning_2022_eana, ghazi_2023_adafest, lim_2024_lazydp}.
%
LazyDP~\cite{lim_2024_lazydp}, the closest to our work, also leveraged the fact that one can defer adding noise until an entry is accessed and add an equivalent, aggregated noise.
However, \cite{lim_2024_lazydp} relies on the fact that the sum of independent Gaussians is also a Gaussian, and it cannot be extended to (non-independent) correlated noises. Our optimization for DLRM is entirely different from LazyDP.
%
The others~\cite{Ning_2022_eana, ghazi_2023_adafest} modify the DP algorithm itself, affecting privacy and accuracy.


\noindent \textbf{Near/in-memory processing.}
Near-memory processing (NMP) runs memory-intensive workloads closer to memory.
%
Prior works explored running compute inside the CXL controller (\emph{e.g.}, for LLM inference~\cite{park_2024_lpddr, gu_2025_cent}, vector database search~\cite{ko2025cosmos, sim2022computational}, and DLRM inference~\cite{ke2020recnmp, yun_2024_clay}), DIMM (\emph{e.g.}, for DLRM~\cite{kwon_2019_tensordimm, park_2021_trim, ke_2022_nmp_axdimm} and database operations~\cite{ke_2022_nmp_axdimm}), network switches~\cite{huo_2024_pifs, huangfu_2022_beacon}, and SSD controllers~\cite{mark_2021_recssd, xuan_2022_rmssd}.
Processing-in-memory (PIM) embeds compute logic directly in the memory hardware to enjoy even higher bandwidth~\cite{park_2024_attacc, yun_2024_duplex, he_2025_papi, lee_2025_paise, lee_2021_hbm_pim}. \sys leverages NMP on a CXL controller for correlated noise generation.




%% file: Chapters/9_discussion.tex
\subsection{Applicability to Large-scale Training}
\label{sec:Discussion}
While our evaluation focuses on single-node configurations (economical setups), the idea of \sys naturally extends to multiple nodes, because correlated noise generation of each model dimension is independent (noise generation for different dimensions can be easily parallelized across nodes). However, if additional nodes provide sufficient aggregate CPU/GPU memory, the relative benefit of the CXL NMP device diminishes. We leave the exploration of these scaled-out setups as future work.

%% file: Chapters/10_conclusion.tex
\section{Conclusion}
\label{sec:Conclusion}

DP training with correlated noise is an emerging technique whose system implications have yet to be thoroughly studied.
We conducted a systematic study of the new technique and found several major bottlenecks when applied to DLRMs and billion-parameter models on an economical setup.
We introduced \sys, a framework for efficient DP training with correlated noise.
When baseline approaches fail to deliver competitive performance, hardware/software designs of \sys can deliver 1.23--10.82$\times$ speedup.

%% file: Chapters/osdi22_ae_appendix_template.tex
\appendix
\section{Artifact Appendix}

\subsection*{Abstract}

Our artifact is available on GitHub.
The repository contains our implementation of Cocoon, a system for differentially private training with correlated noise, along with scripts to reproduce the experiments in this paper.
This codebase extends \texttt{fastDP} with noise distribution,
generation, CPU offload pipelines, and noise preprocessing for embedding.

\subsection*{Scope}

This artifact allows readers to validate the performance claims in the paper.
Figures 3, 5, 14, 15 and 18 are scripted runs, and all other figures can be reproduced with appropriate changes.
The getting-started script checks correctness and throughput on a single GPU in less than 30 minutes.

\subsection*{Contents}
The repository consists of five directories: \texttt{fastDP/} contains the
core library including the engine and noise worker. \texttt{examples/} holds experiment scripts for LLM, DLRM, and image classification. \texttt{benchmark/} provides a getting-started script with correctness checks and throughput benchmarks. \texttt{pfl/} contains BandMF matrix solvers from pfl-research. \texttt{train\_utils/} provides shared training utilities.

\subsection*{Hosting}
The artifact is available on GitHub (\url{https://github.com/psu-paws/Cocoon}), \texttt{osdi26} branch. The \texttt{main} branch of the GitHub repository will be actively updated. 

\subsection*{Requirements}

All main experiments were run on a machine with two sockets of 4$\times$ NVIDIA RTX A5000 24GB GPUs and an Intel Xeon Gold 6330 CPU with 128 GB DRAM.
Parameters must be tuned to match the machine's topology and memory configuration.